\documentclass[12pt]{iopart}
\usepackage{graphicx}
\usepackage{hyperref}
\usepackage{amssymb}

\newtheorem{theorem}{Theorem}[section]
\newtheorem{lemma}[theorem]{Lemma}

\newtheorem{definition}[theorem]{Definition}

\renewcommand{\theequation}{\thesection.\arabic{equation}}

\begin{document}

\title[Complex bifurcations]{Complex bifurcations in B\'enard-Marangoni  convection}

\author{Sergey Vakulenko$^{1,2}$ and Ivan Sudakov$^3$} 

\address{$^1$ Institute for Problems in Mechanical Engineering, Russian Academy of Sciences, St. Petersburg 199178, Russia}
\address{$^2$ Saint Petersburg National Research University of Information Technologies, Mechanics and Optics, St. Petersburg 197101, Russia}
\address{$^3$ Department of Physics, University of Dayton, Dayton, OH 45469-2314 USA}

\ead{isudakov1@udayton.edu}
\vspace{10pt}

\begin{abstract}
We study the dynamics of a system defined by the Navier-Stokes equations for a non-compressible fluid with Marangoni boundary conditions in the two dimensional case. We show that more complicated bifurcations can appear in this system for a certain nonlinear temperature profile as compared to bifurcations in the classical Rayleigh-B\'enard and B\'enard-Marangoni systems with simple linear vertical temperature profiles. In terms of the B\'enard-Marangoni convection, the obtained mathematical results lead to our understanding of complex spatial patterns at a free liquid surface, which can be induced by a complicated profile of temperature or a chemical concentration at that surface. In addition, we discuss some possible applications of the results to turbulence theory and climate science.   

\end{abstract}


\submitto{\jpa}

\maketitle

\section{Introduction}
\label{intro}


The study of bifurcations in fluid and climate systems has attracted the attention of many researchers in connection with climate tipping point problems (see \cite{Lenton} for an overview) and turbulence \cite{Chorin, NRT}. In particular, in \cite{NRT} a hypothesis is pioneered that fluid turbulence can appear as a result of bifurcations from a simple dynamics to a chaotic dynamics.

\begin{figure}[t]
\center
\includegraphics[width=0.9 \linewidth]{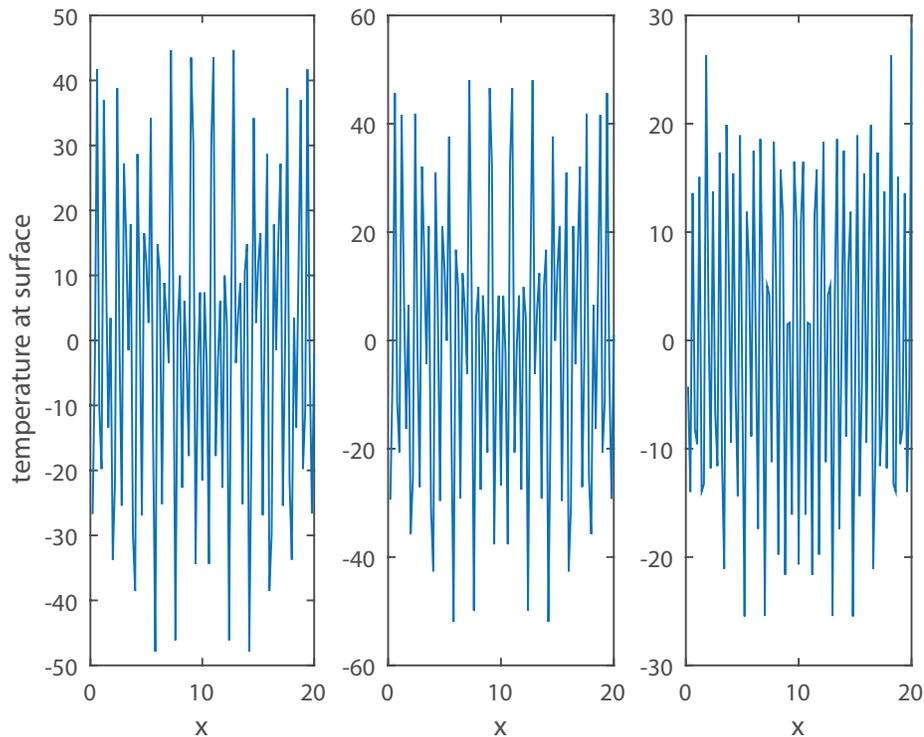}
\caption{Temperature patterns at the surface $y=0$ as functions of $x$ for  times $t=1000, 2000$ and $t=3000$. Here slow dynamics at the bifurcation point is defined by $3$ slow modes. Dynamics of these modes is defined by the Lorenz system. The wave numbers of these modes are $k_1=1, k_2=2.1$ and $k_3=3.3$. 
For generic $k_i$  such patterns are quasiperiodic and essentially one-dimensional since they are localized at the top boundary $y=0$ (rather then the space periodic patterns found for the Rayleigh-B\'enard problem \cite{Ma} and for the Lorenz model)}
\label{Fig00}
\end{figure}

In this paper, we show, in an analytical way, that some spatially inhomogeneous fluid systems are capable of exhibiting a large spectrum of complicated bifurcations. At a bifurcation point, we observe a transition from a steady state attractor to chaotic dynamics and complex spatio-temporal patterns, which are quasiperiodical in space and chaotic in time. These patterns describe coherent turbulent structures
(see Fig. \ref{Fig00}).

We have an explicit analytic description of turbulent onset. In these turbulent patterns, the chaotic attractor dimension and dynamics are controllable by space inhomogeneities, and although that control is complicated, it is quite constructive. The dynamics at the bifurcation point is defined by so-called normal forms. We show that all kinds of structurally stable dynamics can appear as normal forms as we vary system space inhomogeneities. 

To explain the physical ideas behind complicated mathematics  and to understand this new bifurcation mechanism,  let us recall the classical results  on Rayleigh-B\'enard and B\'enard-Marangoni 
bifurcations. They show that, for simple linear temperature profiles  $U(y)$,  where $y$ is the vertical coordinate, the bifurcation is a result of  a single mode 
 instability. This mode is periodic in $x$ with period $T=2\pi/k$,  where $k$ is a wave vector.
The instability arises if, for a given $k$,  the real part  $r_k(b)=Re \lambda_{k}(b)$ of the eigenvalue $\lambda_{k}$  corresponding to this  mode
goes through $0$ as a bifurcation parameter $b$ passes through a critical point $b=b_c$.  
For $b < b_c$ we determine that the trivial zero solution of fluid equations is stable, and for $b$ close to $b_c$ we find  
stable solutions describing periodical patterns.  The amplitudes of these patterns  can be found by a  system of  differential equations \cite{GJ, Ma, Sengul}, which can be considered as a "normal form" of the system at the bifurcation point. That normal form determines the dynamics of slow modes in the system. The dynamics of the  fast modes is captured, for large times, by slow modes \cite{Holmes}. 
For many bifurcations the normal form is defined by a
system with quadratic nonlinearities, since for small amplitudes the main nonlinear contributions are quadratic 
\cite{Holmes}. 
In particular, for the Marangoni case, an analysis of this system  shows that the system bifurcates into two steady state solutions, which are local attractors \cite{Sengul}. 

\begin{figure}[t]
\center
\includegraphics[width=0.9 \linewidth]{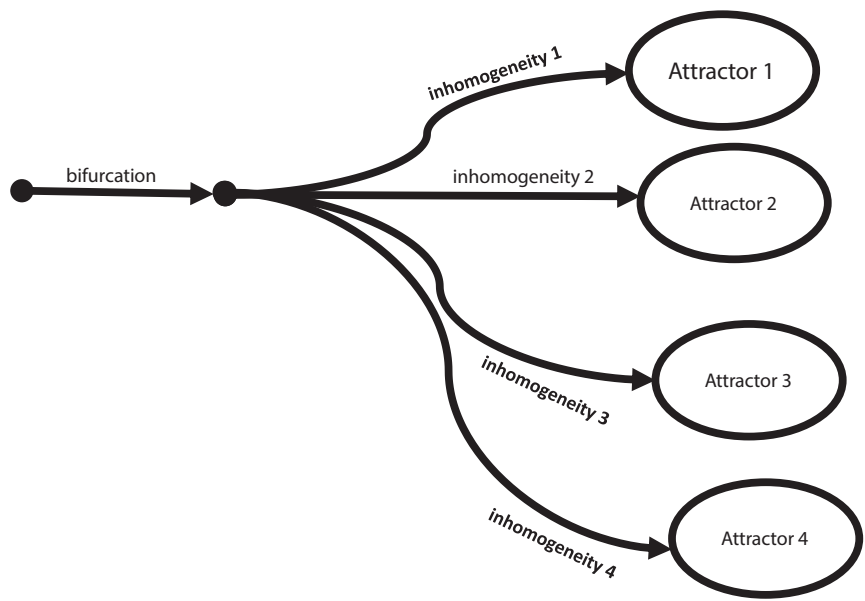}
\caption{Control of bifurcations by  space inhomogeneities in spatially extended systems. 
This scheme works for reaction-diffusion systems \cite{Vakbook} and we propose that it may work for climate systems.}
\label{Pic}
\end{figure}

We use a new scheme  to obtain much more complicated bifurcations and attractors.  This scheme 
is illustrated by Fig. \ref{Pic}. 

\begin{enumerate}
\item First, the
number of slow modes $n$ is controllable by 
the space inhomogeneity.

\item These slow modes are associated with eigenfunctions $\psi_i$ of a linear operator that describe linearization of system.  The functions $\psi_i$   are defined by  different wave vectors $k_i$, $i=1,..., n$. Moreover, the values of $k_i$ are completely controlled by the system space inhomogeneity.  We can get any sets of $k_i$.  

\item Lastly, the quadratic normal form is also completely controllable by the space inhomogeneities. We can obtain any quadratic system by a variation of 
system space  inhomogeneities.  
\end{enumerate}

The key point (iii) allows us to exhibit the existence of chaotic dynamics. This is found in \cite{Stud}, where it is shown that all 
kinds of structurally stable dynamics can be generated by quadratic systems. Thus, all structurally stable dynamics can appear as a result of bifurcations. Such bifurcations can be named {\em superbifurcations}.

Note that space inhomogeneities  consist of two terms. The first term is basic and depends on the vertical coordinate $y$ only. An appropriate  choice  of this term leads to realization of  points 
(i) and (ii).  The second term is small with respect to the first but depends on both vertical and horizontal coordinates.  Variations of this term allow us to control the normal forms. The fact that the normal forms sharply depend  on small inhomogeneities may be interesting 
for climate theory.  Indeed, we need points (i) and (ii)  to create a number of slow modes whereas
a possibility of complex interactions between these modes  follows from  (iii).  A typical climate system involves  a number of slow modes    
(see \cite{Ma, Ashwin}). We can expect that in such systems small space inhomogeneities can 
lead to complex bifurcations. This property  may be important for tipping point theory, which, up to now, considered relatively simple bifurcations in simplified models (see \cite{Ashwin} for an overview).

In this paper, we consider local bifurcations only where the dynamics at the bifurcation point is weakly nonlinear.  Although we show the existence of complex phenomena and the appearance of chaotic attractors of all dimensions, fully nonlinear systems can also exhibit non-local bifurcations that are harder to describe 
by the analytical methods of this paper. However, one can expect that if even local bifurcations lead to all kinds of chaos of any dimension, then the same fact also holds for non-local ones.

The solution methodology can be described as follows. 
According to standard bifurcation ideas, at the bifurcation point the system of solutions can be represented as  sums of contributions of slow modes. Each contribution is defined by the corresponding magnitude $X_k(t)$, which is a slow function of time $t$. The spatial patterns corresponding to our solutions are complex, they are quasiperiodic in $x$ (horizontal axis) and localized at the surface. Dynamics of the magnitude $X_k$ may be time periodic or even chaotic. We thus have complex coherent spatio-temporal patterns quasiperiodic in space, localized at the surface, and evolving in time in a complex  manner (see Fig. \ref{Fig00}). 
Moreover, we can control the pattern structure and  dynamics.
The mathematical realization of that control is based on a vector-field realization approach (the RVF method, see  section \ref{RVF} in the Appendix and \cite{Stud,Arch,Pol2,Pol3}). This method was successfully applied to many dissipative systems of chemical kinetics and neural network theory. Here, we first use the RVF for fluid dynamics.

We believe that these ideas can be used for many systems, for example, for climate systems. Indeed, climate systems include fast and slow variables and involves spatial inhomogeneities \cite{Stm}. One can expect that the mechanism of generation of complex large time behaviour, outlined above, works in these  systems.

\begin{figure}[t]
\center
\includegraphics[width=0.9 \linewidth]{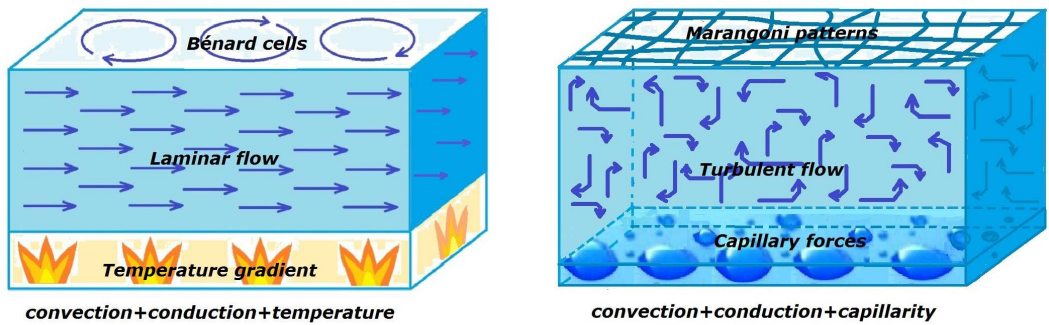}
\caption{On the left: B\'enard convection cells, which are formed by temperature gradients; on the right: the Marangoni patterns are driven by surface tension gradients.}
\label{Fig0}
\end{figure}

In this paper, for simplicity, we study a toy model having physical applicability.
We consider the Navier-Stokes (NS) equations  for a non-compressible fluid
  with the Marangoni boundary conditions in the two dimensional case. These equations
 describe hydrodynamical systems involving convection, heat transfer
and capillarity, and exhibit interesting pattern formation effects \cite{Thomson} (for example,
B\'enard cells studied in many works
\cite{GJ, Sengul, Bracard} and references therein). As indicated in Fig.\ref{Fig0},
the Marangoni flows are driven by surface tension gradients. In general, surface tension  depends
on both the temperature and chemical composition at an interface.  Therefore, these
flows may be generated by gradients in either temperature or chemical concentrations.

In physical and chemical systems the Marangoni effects are well-known and this may have potential applications for geosystems.
For example, shallow permafrost lakes in tundra emit a huge amount of methane that affects the atmospheric dynamics and atmospheric thermodynamics. Apparently, the "surface tension" in the system of permafrost lake patterns on the tundra's surface will define methane emission regimes \cite{Sudakov}. However, the Marangoni effects also can influence large ecosystems, for example oceanic ones, if there are salinity gradients \cite{Sal}. In addition, the surface tension of green algal which bloom on the surface of big lakes in Southeast Asia (and fully cover the lake surface), changes the internal physical and chemical properties of these lakes \cite{Botte}. Finally, the surface tension at the air-water-sedimentary rock interface of geothermal hot springs defines  the dynamics of water flows \cite{Gold}.

An interesting situation, where these effects are essential, arises when there exist surfactants on the water surface. The capillary forces are also important  for a correct description of air bubble motion. In this paper, we consider thermocapillary Marangoni effects; however, the same analysis is valid when these effects are induced by convection, capillary forces and diffusion, for example, 
by salinity gradients.

Our plan to find  complicated bifurcations is as  follows.  
Let us first fix  the fluid viscosity and all other parameters except the Marangoni number $b$, which serves as a bifurcation parameter. This Marangoni parameter   $b$ is  the
ratio of the destabilizing surface tension gradient to the 
forces generated by thermal and viscous diffusion \cite{Sengul}.
We consider general nonlinear profiles  $U(y)$ depending on the vertical coordinate only. First, we remove  nonlinear terms and consider the corresponding linearized problem.  
As above, the eigenfunctions of this problem  are periodic in $x$ with the period $T=2\pi/k$, where $k$ is a wave vector.  For example,  the fluid stream function $\psi(x,y)$ induced by these modes has
 the form $\psi_k=\Psi_k(y, \lambda) \cos(kx)$. For large  Prandtl numbers 
 we  obtain an asymptotic equation for $\lambda$. 
 The physical meaning of this equation is transparent; for each wave vector  the eigenvalue $\lambda(k)$ is defined by  an interaction between the corresponding mode 
 $\psi_k$ and the inhomogeneity $U(y)$. Our goal is to find profiles $U(y)$ for which the stability loss occurs simultaneously for many $k$, i.e., we have  that the real parts $r_k(b)=Re \lambda_{k}(b)$ pass through $0$ for all $k=k_1, k_2, ..., k_N$ as
$b$ goes through $b_c$.

The key idea in finding such profiles  is as follows.  First we find a profile $V(y)$ such that the spectrum $\lambda(k)$ is independent of $k$ within a large interval of 
values $k$.   For example, we  take  $V$
   as a step function, where a step localized at the surface $y=0$, for example, at
 $y=z_0 << h$, where $h$ is the depth of fluid layer. There are other possible kinds of $V$, for example, exponential profiles.
  Then the interaction between the fluid  and the  inhomogeneity  $U$ weakly depends on the wave vector $k$. Therefore, the stability loss
 will occur simultaneously for a number of $k$ (see comment at the formulation of Theorem  \ref{6.2}).     
Indeed,   numerical simulations and analytic  arguments (see Sect. \ref{sec4} and Fig. \ref{Fig1})  show that, for an appropriate $b=b_c$,  the quantities $r_k(b)$ changes their sign at $b=b_c$ for 
a set of $k$. Then we take the profile $U$ as a weak perturbation of $V$.  This trick aims to attain two key goals: 

{\bf a})  we obtain the stability loss at prescribed $k_1, k_2, ..., k_N$ for each $N$ and these $k_j$ define wave vectors of slow modes; {\bf b}) we shift all the other eigenvalues corresponding
to $k \ne k_j$ towards the left half-plane  $Re \lambda  < 0$ and therefore the corresponding modes are fast.  

Furthermore, we take into account nonlinear terms. At the bifurcation point, by standard methods of invariant manifold theory \cite{Pol2, Pol3,He}, we derive a system of differential equations with 
 quadratic nonlinearities for  mode magnitudes $X_j(t)$, $j=1,..., N$, 
that is reminiscent of the Lorenz system. Previous results \cite{Stud, Arch} show that the corresponding dynamical system  can have periodic or  chaotic trajectories.

 Note that the main technical mathematical difficulties are connected with points (i) and (ii).  The application the method of realization of vector fields (see Appencix and  \cite{Vakbook, Pol2, Pol3})
 of (iii) is standard.


These mathematical results admit a simple physical interpretation. They describe a complex spatial pattern at a free water surface, which can be induced by 
a complicated profile of temperature or a chemical concentration at that surface.  Note that such profiles of salinity concentrations  arise in real fluid systems 
 (see \cite{Sal}).  To obtain the complicated profile, we use spatially distributed inhomogeneous sources; however,
an analytical proof of the existence of even more complex solutions determined by a non-trivial temperature profile is obtained 
in paper \cite{Aristov} for the case of a planar free B\'enard-Marangoni convection.

This paper is organized as follows.
In the next section we formulate the Marangoni problem. In Section \ref{sec3} we introduce the main linear operator associated 
with the problem. 
 In Section \ref{sec4} we investigate the spectrum of this operator, and we state the main new conclusion of the paper and its proof.    
In section \ref{sec5} the Marangoni initial boundary value problem is reduced to a system of differential equations with quadratic nonlinearities describing the dynamics of
the main modes at the bifurcation point (it follows the standard technique, see \cite{Sengul}).  
Finally, in Section \ref{sec7} we show that quadratic systems obtained in Section \ref{sec5} can exhibit a chaotic large time behaviour.  

The appendix contains all complex mathematical proofs. In particular, 
section \ref{RVF} considers the RVF method introduced by P. Pol\'a\v cik \cite{Pol2,  Pol3},  which is an important tool for analytically proving the existence of chaos.

\section{Marangoni problem for Navier-Stokes equations}
\label{sec2}

We consider
the Navier-Stokes system for an ideal incompressible fluid
\begin{equation}
  {\bf v}_t +  ({\bf v} \cdot \nabla) {\bf v} =\nu \Delta {\bf v} - \nabla p,
\label{OB1}
\end{equation}
\begin{equation}
     \nabla \cdot {\bf v} =0,
\label{div}
\end{equation}
\begin{equation}
   u_t +  ({\bf v} \cdot \nabla) u = D \Delta u + \eta,
\label{OB2}
\end{equation}
  where  ${\bf v}=(v_1(x, y, t), v_2(x, y, t))^{tr}$,
$u=u(x,y,t), p=p(x,y, t)$ are   unknown functions
defined on $\Omega \times \{ t \ge 0 \}$,
 $\Omega$ is the strip
$(-\infty, \infty) \times [0,h] \subset {\bf R^2}$.
Here $\bf v$ is the fluid velocity, where $v_1$ and $v_2$ are the normal and tangent velocity components,
                           $\nu$ and $D$ are 
the viscosity and thermal diffusivity  coefficients, respectively,  $p$ is the pressure,
$u$ is the temperature,  and
$\eta(x,y)$ is a  function describing a distributed
heat source.
By ${\bf v} \cdot \nabla$ we denote the advection operator $v_1 \frac{\partial}{\partial x} +
v_2 \frac{\partial}{\partial y}$. 
The initial conditions are
\begin{equation}
  {\bf v}(x, y, 0)={\bf v}^0(x, y),
\quad { p}(x, y, 0)={ p}^0(x, y),
\quad u(x,y,0)=u^0(x,y).
\label{inidata}
\end{equation}
Let us
suppose that  the unknown functions
are $2\pi$-periodic  in
$x$:
\begin{equation}
  {\bf v}(x, y, t)={\bf v}(x+ 2\pi, y, t),
\quad { p}(x, y, t)={ p}(x+ 2\pi, y, t),
\label{period1}
\end{equation}
\begin{equation}
 u(x, y, t)= u(x+ 2\pi, y, t),
\label{period2}
\end{equation}
and that  $u^0, p^0, {\bf v}^0$ also are  $2\pi$ -periodic in $x$.
The function $u$ satisfies the Neumann boundary conditions:
\begin{equation}
  u_y(x, y, t)\vert_{y=h} =0, \quad u_y(x, y, t)\vert_{y=0} =0.
\label{boundNeum}
\end{equation}
We assume that the surface $y=h$ is free:
\begin{equation}
  v_2(x, h, t) =0,  \quad  \frac{\partial v_1(x, y, t)}{\partial y}\vert_{y=h}=0.
\label{bound5}
\end{equation}
The Marangoni boundary condition at $y=0$ is defined by a relation
connecting the tangent velocity component and the tangent gradient of the temperature:
\begin{equation}
 { v_1}_y (x,y,t)\vert_{y=0}   = - b u_x(x,0,t),
\label{Maran1}
\end{equation}
where $b >0$ is the Marangoni parameter.   For $v_2$ at $y=0$ one has
\begin{equation}
 { v_2} (x,0,t)   = 0.
\label{Maran1h}
\end{equation}
Let us assume that
\begin{equation}
  \langle  \eta, 1 \rangle=\int_{\Omega} \eta(x,y) dxdy =0,
\label{aver1}
\end{equation}
where $\langle u, v \rangle $ is the scalar product in $L_2(\Omega)$:
\begin{equation}
 \langle u, v \rangle =\int_0^h \int_0^{2\pi} u(x,y) v(x,y) dx dy.
\label{inprod}
\end{equation}
Note that if $u(x,y, t)$ is a solution to (\ref{OB2}),(\ref{boundNeum}) and (\ref{period2}),  then for  any constant $C$ the function
$u(x,y,t)  + C$ also is a solution.

We use below
the stream function - vorticity   formulation of these
equations in order to exclude the pressure $p$.
Introducing the vorticity $\omega$ and the stream function $\psi$, we obtain \cite{Chorin}
\begin{equation}
\Delta \psi=-\omega,
\label{OBEstream2}
\end{equation}
where the velocity  $\bf v$ can be expressed
 via the stream
function $\psi(x,y)$  by the relations
$v_1=\psi_y, v_2=-\psi_x$.
 Equations (\ref{OB1}), (\ref{div})
and (\ref{OB2})  take the form  \cite{Chorin}
\begin{equation}
\omega_t + \{\psi, \omega\} =\nu \Delta \omega,
\label{OBEstream1}
\end{equation}
where  $\{\psi, \omega \}=
  \psi_y \omega_x -\psi_x \omega_y$,
\begin{equation}
u_t + \{ \psi, u \} =D\Delta u + \eta.
\label{heat1}
\end{equation}
The boundary conditions  become
\begin{equation}
  \psi(x,y,t)=\psi(x+2\pi,y,t),  \quad
\omega(x,y,t)=\omega(x+2\pi,y,t),
\label{boundstream1}
\end{equation}
\begin{equation}
 {\psi} (x,y,t)\vert_{y=h}   = \omega(x, y, t)\vert_{y=h}=0,
\label{boundstream2}
\end{equation}
\begin{equation}
 {\psi} (x,y,t)\vert_{y=0} =0, \quad   \omega(x, y, t)\vert_{y=0}= bu_x(x,0,t),
\label{Maran}
\end{equation}
\begin{equation}
  u_y(x, y, t)\vert_{y=h} =0, \quad u_y(x, y, t)\vert_{y=0} =0.
\label{boundNeum2}
\end{equation}
We set initial conditions
\begin{equation}
  u(x,y,0)=u_0(x,y),  \quad \omega(x,y, 0)=\omega_0(x,y,t).
\label{Initdata}
\end{equation}

Global  existence and uniqueness of the solutions to the Marangoni initial boundary value problem (IBVP) (\ref{OBEstream2})-(\ref{Initdata})  on $(0, +\infty)$ follows from results in \cite{Sengul}. See also
\cite{Pardo} for the stationary 3D case.

\section{Linearized problem}
\label{sec3}

We follow the classical approach developed for the Rayleigh-B\'enard and B\'enard-Marangoni convection
\cite{GJ, Ma, Sengul, Bracard}. Assume that
the temperature field $u$ is a small $\gamma$ -perturbation of a vertical profile $U(y)$. Here  $\gamma >0 $ is a small parameter
independent of the viscosity $\nu$ (this assumption is  important): $\gamma < \gamma_0(\nu)$.
Let 
$U(y)$ be a $C^{\infty}$ smooth function of $y \in [0,1]$  and 
$u_1(x,y)$ is another smooth function  $2\pi$ -periodic in $x$.  We assume that
\begin{equation}  \label{suppU}
U(y) =0,  \quad \forall y \in [0,\delta_1),
\end{equation}
for a $\delta_1  \in (0, h)$ and that the support of $u$ does not intersect  the boundary $y=0$:
\begin{equation}
 u_1(x,y)=0   \quad \forall  y \in [0,\delta_1) \quad \forall x \in {\bf R}.
\label{suppu1}
\end{equation}

We set
\begin{equation} \label{zamena}
 u=U + \gamma u_1 + \gamma w,  \quad  \psi=\gamma \tilde \psi, \quad \omega=\gamma \tilde \omega,  
\end{equation}
where $\tilde \omega(x,y, t), w(x,y,t)$ and $\tilde \psi (x,y, t)$ are new unknown functions. 
 Substituting (\ref{zamena}) 
 into (\ref{OBEstream1}), (\ref{heat1}), one obtains
\begin{equation}
 \tilde \omega_t=\nu \Delta \tilde \omega  - \gamma  \{\tilde \psi,  \tilde \omega \},
\label{eveq10}
\end{equation}
\begin{equation}
w_t= D \Delta w   - \{\tilde \psi, U \} - \gamma (\{\tilde \psi, u_1 \} + \{\tilde \psi, w \} + \eta_1),
\label{eveq11}
\end{equation}
where  $\eta_1=\gamma^{-1} (\eta +  D U_{yy} +D \Delta u_1)$. We  assume that $\eta_1$ is a smooth bounded function, $\sup |\eta_1| < C$,
where $C$ does not depend on $\gamma$.
Note that the space inhomogeneous source $\eta$ plays an important role in the bifurcation construction. This allows us to create a nontrivial non-perturbed temperature (salinity) profile
$U(y)$ close to a step-function. Such complicated profiles can appear in fluid systems \cite{Sal}.

\subsection{ Function spaces}
\label{sec: 7.1}

We  use standard Hilbert  spaces \cite{He}.
 We denote by $H=L_2(\Omega)$ the Hilbert space  of measurable, $2\pi$- periodical in $x$ functions defined
on $\Omega$
 with bounded  norms $|| \ ||$, where $||u||^2=\langle u, u\rangle$ and $\langle, \rangle$ is
the inner product defined by (\ref{inprod}).
Let us  denote by  $H_{\alpha}$ the fractional spaces
\begin{equation}
H_{\alpha}= \{ \omega :  ||\omega||_{\alpha} =||(I-\Delta_D)^{\alpha} \omega|| < \infty \},
\label{Hal}
\end{equation}
here $\Delta_D$ is the Laplace operator  with the standard domain corresponding to the zero Dirichlet boundary conditions:
\begin{equation}
Dom \ \Delta_D=\{ \omega: \omega \in W_{2,2}(\Omega), \quad \omega(x, y)\vert_{y=0, y=h}=0 \},
\label{DomDelta}
\end{equation}
here $W_{q,2}(\Omega)$ denote the standard Sobolev spaces.
 Let  $\tilde H_{\alpha}$  be another fractional space associated with $L_2(\Omega)$:
\begin{equation}
\tilde H_{\alpha}= \{ u :  ||u||_{\alpha} =||(I-\Delta_N)^{\alpha} u|| < \infty \},
\label{Bp}
\end{equation}
where $\Delta_N$ is the Laplace operator with the  domain corresponding to the zero Neumann boundary conditions
\begin{equation}
Dom \ \Delta_N=\{u: u \in W_{4,2}(\Omega), \quad u_y(x, y)\vert_{y=0, y=h}=0 \}.
\label{DomDeltaN}
\end{equation}
Below we sometimes omit the indices $N, D$.  This choice of the domain for $\Delta_N$ is connected with
a special choice of the main function space for the $u$ -component, which should be a more regular one than the $\omega$ -component.
We choose ${\mathcal H}=H \times \tilde H_1$ as a phase space for IBVP  (\ref{OBEstream2}) -(\ref{Initdata}).

\subsection{Linear operator $L$ and existence of solutions}

Removing the  terms  of  order $\gamma$ in  (\ref{eveq10}),  (\ref{eveq11}) we obtain a linear evolution equation associated with a linear operator $L$.
The spectral problem for this operator $L$  is defined by the following equations:
(we omit the tilde in notation for $\tilde \psi, \tilde \omega$):
\begin{equation}
\lambda \omega=\nu \Delta \omega,
\label {Sp1}
\end{equation}
\begin{equation}
\lambda w=D \Delta w +  \psi_x U_y,
\label {Sp2}
\end{equation}
where the functions $w$ and $\omega$ satisfy the boundary conditions
\begin{equation}
   \omega(x,h)=0,  \quad  \omega(x,0)= \mu w_x.
\label {Sp3}
\end{equation}
\begin{equation}
w_y(x, y)\vert_{y=0, h}=0.
\label {Sp4}
\end{equation}
Moreover, 
\begin{equation}
\Delta \psi=-\omega, \quad  \psi(x,0)=\psi(x, h)=0.
\label {Sp5}
\end{equation}
   This spectral problem  is investigated in the next sections but first we consider the general properties of $L$.
This operator has a dense domain  $ Dom \ \Delta_D \times Dom \Delta_N$ in the Hilbert phase space ${\mathcal H}=H \times \tilde H_1$.  
The operator $-L$ is sectorial as can be proved by  Theorem 1.3.2 from \cite{He} (see \cite{Arch}).  Furthermore, one can  show that the operator $L$ has a compact resolvent, and therefore, the spectrum of $L$ is discrete  \cite{Arch}.
In the coming section we study the spectrum of  the operator $L$.

The fact that the linear operator $L$ is sectorial and some  results on smoothness of nonlinear terms in equations (\ref{eveq10}), (\ref{eveq11})  
\cite{Arch} show that 
IBVP   (\ref{OBEstream2}) -(\ref{Initdata}) defines a $C^1$-smooth local semiflow in ${\mathcal H}$.
For results on global existence see \cite{Sengul}, where the 2D non-stationary case is considered, and \cite{Pardo} for
the 3D-stationary case.

\section{ Spectrum of operator $L$  }
\label{sec4}

\subsection{ Some preliminaries
  }\label{sec:8.1}

Let us consider the spectral problem in (\ref{Sp1}), (\ref{Sp2}) and (\ref{Sp3}).  
For any $U(y)$ this problem has the trivial eigenfunction  $e_0=(0,1)^{tr}$, where $\omega=0, w=1$,  with
the zero eigenvalue $\lambda$.
We consider eigenfunctions $e(x,y,\lambda)$ with eigenvalues $\lambda \in {\bf  C}_{1/2}$, where 
 ${\bf C}_a$ denotes the half-plane 
\begin{equation}
{\bf C}_a= \{ \lambda \in {\bf C}:   Re \ \lambda > -a \}.
\label{Ck}
\end{equation}
Since $U$  depends only on $y$, 
we  seek the eigenfunctions  of the form 
\begin{equation}
   \quad \quad w(x,y, \lambda)= w_k(y, \lambda) \exp(ikx),
\label{wF}
\end{equation}
  \begin{equation}
   \psi(x,y, \lambda)= \psi_k(y, \lambda)  \exp(ikx), \quad   \omega(x,y, \lambda)= \omega_k(y,\lambda)  \exp(ikx).
\label{psiF}
\end{equation}
For  $\omega_k$, $\psi_k$ and $w_k$ one obtains the following
boundary value problem:
\begin{equation} \label {BVK1}
\frac{\partial^2 \omega_{k}}{\partial y^2} - k_{\nu}^2 \omega_k=0,   \quad  \omega_k(h,\lambda)=0, \quad \omega_k(0, \lambda)= i b k w_k(0, \lambda),
\end{equation}
where $k_{\nu}^2=k^2 +\lambda/\nu$,
\begin{equation} \label {BVK2}
\frac{\partial^2 \psi_{k}}{\partial y^2} - k^2 \psi_k=-\omega_k,   \quad  \psi_k(h, \lambda)=0, \quad \psi_k(0, \lambda)=0,
\end{equation}
\begin{equation} \label {BVK3}
\frac{\partial^2 w_{k}}{\partial y^2}  - \bar k^2 w_k=D^{-1} ik U_y(y) \psi_k,   \quad  \frac{\partial w_k(y, \lambda)}{\partial y}\vert_{y=0, h}=0,
\end{equation}
where $\bar k=\sqrt{k^2 +\lambda/D}$.
   Note that $w_{-k}$ are functions, complex conjugate to $w_k$
and $\bar k$ and $k$ are involved in the above equations only via $\bar k^2$ and $k^2$, respectively. Therefore, we can suppose, without loss of generality, that $k > 0$, and $Re \  \bar k >0$ for $\lambda \in {\bf C}_{1/2}$. We also assume that
\begin{equation} \label{hviscos}
h=10 \log \nu, \quad  \nu >> 1.
\end{equation}

The solution of problem in (\ref{BVK1})-(\ref{BVK2}) is defined by
 \begin{equation}
   \omega_k(y,\lambda) = \beta_k \frac{\sinh (\bar k_{\nu}(h-y))}{\sinh(\bar k_{\nu} h)},    \quad  \beta_k(\lambda)=  i b k w_k(0, \lambda) ,
\label{omFk}
\end{equation}
\begin{equation}
 \psi_k(y, \lambda)=- \nu \beta_k \lambda^{-1} \Phi_k(y, \lambda),
\label{psiF2}
\end{equation}
where
\begin{equation}
 \Phi_k(y, \lambda)=\frac{\sinh(k h) \sinh(\bar k_{\nu}(h-y)) -  \sinh(\bar k_{\nu}h)  \sinh (k(h-y)) }{ \sinh(\bar k_{\nu} h) \sinh(kh)}.
\label{psiFi}
\end{equation}
Note that relation (\ref{psiF2}) is correctly defined for all $\lambda \in {\bf C}_{1/2}$,  in particular, for  $\lambda=0$. Indeed,   for     
small $\lambda$ 
\begin{equation}
\bar k_{\nu}-k=\sqrt{k^2 + \lambda \nu^{-1}} -k=  \lambda (2\nu k)^{-1} +  O(\lambda^2\nu^{-2} k^{-3})
\label{dk}
\end{equation}
that gives 
\begin{equation}
   \psi_k(y, \lambda)=\beta_k(\lambda) \frac{ y\sinh(kh) \cosh(k(h-y)) - h \sinh(ky) }{2k\sinh^2( k h)}  + \phi(y, k),
\label{psiF4}
\end{equation}
where
$$
|\phi(y, k, \lambda)| <  c|w_k(0, \lambda)||\lambda| (k\nu)^{-1}, \quad 0 < y < h.
$$
 For large $\nu_0$  and $|\lambda| << \nu$ assumptions  (\ref{hviscos})  allow us to simplify (\ref{omFk}) and (\ref{psiF2}). By (\ref{omFk})
 we obtain 
\begin{equation} \label{omegaF}
  \omega_k(y, \lambda)= \beta_k(\lambda)  (\exp(- k y)  +  \tilde \omega_k(y, \nu)),
\end{equation}
where for each $s \in (0,1)$ and $|\lambda| < \nu^s$
\begin{equation} \label{tildeomega}
|\tilde \omega_k(y, \nu)| < C_s (|\lambda| (k \nu)^{-1} \exp(- ky) +  \exp(-k h)), \quad y \in [0,h],  
\end{equation}
where $C_s>0$ are constants independent of $s$ and $k$.
This estimate and (\ref{psiF4}) give
\begin{equation} 
\label{psiFa}
\psi_{k}(y, \lambda)=\beta_k(\lambda)(\bar \psi_k(y, \lambda) + \tilde \xi_k(y,  \lambda)),
\end{equation}
where 
\begin{equation} 
\label{psiFa1}
\bar \psi_k(y, \lambda)=\frac{y}{2k} \exp(- k y) ,
\end{equation}
and for $|\lambda| < \nu^s$
\begin{equation} \label{tildepsi}
|\tilde \xi_k(y,  \lambda)| <  \bar C_s (|\lambda| k^{-1} \nu^{-1} \exp(-ky) +    \exp(-k h)),   \quad y \in (0,h),
\end{equation}
 where constants $\bar C_s>0$ are uniform in $k, \nu$.

To investigate (\ref{BVK3}), we apply a  lemma.

\begin{lemma} \label{6.1}
{ Let us consider the  boundary value problem on $[0, h]$ defined by
\begin{equation}
w_{yy} - \bar k^2 w= f(y), \quad y \in [0, h],
\label{kwd1}
\end{equation}
\begin{equation}
w_{y}(y)\vert_{y=0, h} =0.
\label{kwb}
\end{equation}
Then
\begin{equation}
w(0)=-\int_0^h f(y)  \rho_{\bar k}(y) dy,
\label{kwc}
\end{equation}
where
\begin{equation}
  \rho_{\bar k}(y) = \frac{\cosh(\bar k(h- y))}{{\bar k} \sinh {\bar k} h}.
\label{kwr}
\end{equation}
}
\end{lemma}

To prove this lemma, we multiply both the right hand and the left hand sides
of Eq.(\ref{kwd1}) by $\rho_{\bar k}$ and integrate by parts
in the left hand side 
$\square$.

Note that  
\begin{equation} \label{rhoest1}
|\rho_{\bar k}(y)- \bar \rho_{\bar k}(y)| < \bar k^{-1} \exp(- \bar k h), \quad  \bar \rho_{\bar k}(y) =\bar k^{-1}\exp(-\bar k y). 
\end{equation}

\subsection{ Main result on spectrum}
\label{sec: 8.2}

The following assertion is a mathematical formalization of the key ideas {\bf i} and {\bf ii} (see the introduction). We show how  one can control the spectrum  of the operator $L$ by the bifurcation parameter  $b$ and space inhomogeneity. Informally, we can obtain any prescribed number of zero eigenvalues with
any prescribed wave numbers.  

\begin{theorem} \label{6.2}
{Let  assumptions (\ref{hviscos}) hold,
 $N$ be a positive integer  and $K_N=\{k_1, ..., k_N \}$ be a subset of ${\bf Z}_+$.
Then 
there exists an open non-empty  interval $J=(b_1, b_2)$, a number $b_c \in J $  and a  $C^{\infty}$ smooth  function $U(y)=U_{K_N}(y, \nu)$ satisfying (\ref{suppU})  and such that  
 for sufficiently large $\nu > \nu_0(K_N)$
the eigenfunctions $\lambda(k, \mu, \nu)$ of boundary value problem (\ref{BVK1}), (\ref{BVK2}) and (\ref{BVK3}) satisfy

({\bf i} )  
for $k \in K_N$   
\begin{equation}
\lambda(k, b_c, \nu)=0,    \quad \lambda(k, b, \nu) < 0,  \   b  \in (b_1, b_c), 
\label{Spec0a}
\end{equation}
and
\begin{equation}
   \lambda(k, b, \nu) > 0,  \ b \in (b_c, b_2),    
\label{Spec0b}
\end{equation}

({\bf ii})   for $k \notin K_N$ 
\begin{equation}
  Re \ \lambda(k, b, \nu) < - \delta_N  \quad  b \in (b_1, b_2)
  \label{Spec}
\end{equation}
where a positive $\delta_N$ is uniform in $\nu$.
}
\end{theorem}

\begin{figure}[t]
\center
\includegraphics[width=0.9 \linewidth]{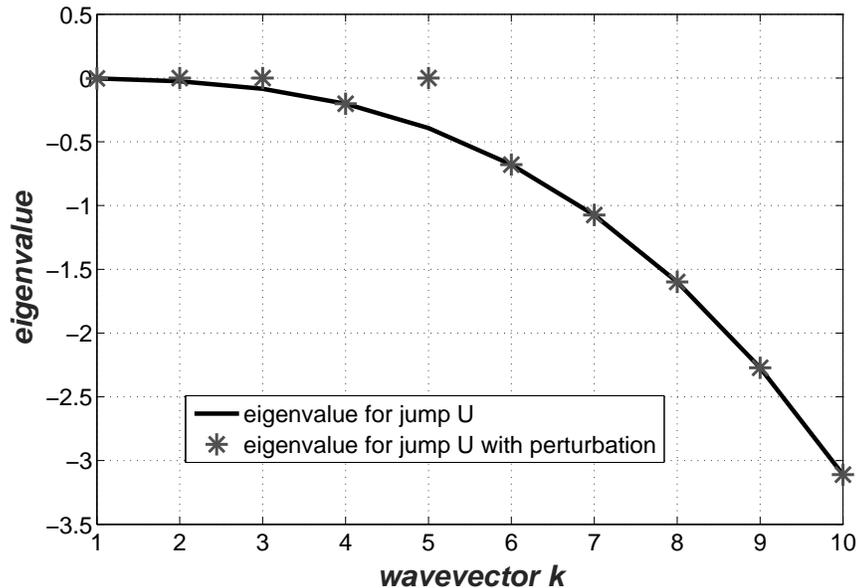}
\caption{This plot show the roots $\lambda$ of the equation for the profile $U$ close to a step function.  The continuous curve shows $\lambda(k)$ for $U=H(y-z0)$ where $z_0=0.0008$.  The starred curve shows $\lambda(k)$ for a specially perturbed $U$,
where $k_1=1,k_2=2, k_3=3, k_4=5$.}
\label{Fig1}
\end{figure}

{\em Ideas behind the formal proof}. The next proof is long, but it is based on 
a simple idea, which we explain informally here. We choose $U(y)$ close to a step-function  $H(y-z_0)$, where $z_0$ is small.  For $U=H(y-z_0)$ and as $\nu \to +\infty$
at the bifurcation point $b=b_c$ the equation  for the eigenvalues $\lambda$ takes the form
\begin{equation}
\exp(-p z_0)= p/k-1,
\end{equation}   
where $p=k+ \sqrt{k^2 +\lambda}$, $k >0$. For $Re \ p > 1/2$ and small $z_0$
this equation has a single root  $p_*(k)$.  This root is  close  to $2k$ for bounded $k$
and the corresponding $\lambda(k)$ is negative and  close to $0$.  If we add a specially adjusted perturbation to
$U$, we can shift $\lambda(k)$ to zero, and obtain $\lambda(k)=0$ for $k=k_1, k_2,..., k_N$.  Fig. \ref{Fig1} illustrates this situation (solutions $\lambda$ are found numerically). 
Note that we also can choose $U$  close to a sharply decreasing function, for example, $U \approx const \exp(-b y),  \ b >>1$. i.e., we have a large class of profiles
leading to superbifurcations.  

The proof can be found in the Appendix. In the coming subsection we consider 
eigenfunctions of the operator $L$ and the conjugate operator involved in the normal form.

\subsection{Eigenfunctions of $L$ with zero eigenvalues}
\label{sec: 8.4}

Let us consider the eigenfunctions $e_k$ of $L$ with the zero eigenvalues. We have $2N+1$ eigenfunctions  including  the trivial one 
$e_0=(0, 1)$.  All the other eigenfunctions have the form
\begin{equation}
e_k(x,y)=\exp(ikx)(\omega_k(y), \theta_k(y))^{tr},  
\label{eig1}
\end{equation}
\begin{equation}
e_{-k}=\exp(-ikx)(\omega_{-k}(y), \theta_{-k}(y))^{tr}, 
\label{eig1a}
\end{equation}
where $k\in \{k_1, k_2, ..., k_N\}$ and 
\begin{equation}
\omega_k= i k A_k  \frac{\sinh (k(h-y))}{\sinh(k h)},  \quad i=\sqrt{-1},
\label{eig2}
\end{equation}
\begin{equation}
\theta_k= A_k \Theta_{k}(y),
\label{eig3}
\end{equation}
where $A_k(\nu)$ are  constants.

To obtain real value eigenfunctions, we  take real and imaginary parts of these complex eigenfunctions.
The real parts of the eigenfunctions, where $ \omega_k, \theta_k$ are proportional to $\sin(k x), \ \cos(k x)$ respectively,
 are denoted by the upper index $+$,  and the imaginary parts, where $ \omega_k, \theta_k$ are proportional to $\cos(k x), \ \sin(k x)$,
are denoted by the upper index $-$.  The  real eigenfunctions of $L$ have the form

\begin{equation}
e_k^+=(\omega_k(y) \sin(kx), \theta_k(y) \cos(kx))^{tr}, 
\end{equation}
\begin{equation}
 e_k^-=(-\omega_k(y) \cos(kx) , \theta_k(y) \sin(kx))^{tr}.
\label{eigpm}
\end{equation}
 Respectively, the  real eigenfunctions of formally conjugate operator $L^*$ are
\begin{equation}
\tilde e_k^+=(\tilde \omega_k(y) \sin(kx), \tilde \theta_k(y) \cos(kx))^{tr}, 
\end{equation}
\begin{equation}
 \tilde  e_k^-=(-\tilde \omega_k(y) \cos(kx) ,\tilde  \theta_k(y) \sin(kx))^{tr}.
\label{eigpms}
\end{equation}
We  have the  relations
\begin{equation}
\tilde\theta_k^{+}=a_k  \cosh( k(h-y)), \quad
\label{eigplus}
\end{equation}
where $a_k$ are coefficients.  One can show that there exist no generalized eigenfunctions of the operator $L$ \cite{Arch}.

In the remaining part of this paper we describe a formal mathematical realization of the key point {\bf iii}, i.e. a control of the normal form that determine the dynamics at the bifurcation point. That control is based on the method of realization of vector fields (RVF) (see Appendix).


\section{Normal form at the bifurcation point}
\label{sec5}

 Assume that
$\gamma >0$ is a small parameter, and $|b-b_c| < \gamma^2$, i.e., we are seeking solutions at the bifurcation point.
 In this section, we reduce the Navier-Stokes  dynamics to a system of ordinary differential equations. This reduction is standard and follows from known works
\cite{Sengul}. 

Let  $E_N$ be
the finite dimensional subspace
$E_N=Span \{e_0, e_1^+,  ..., e_{N}^+, e_1^-, ..., e_N^{-} \}$
of the phase space ${\mathcal H}$, where $e_j^{\pm}=(\omega_j^{\pm}, \theta_j^{\pm})^{tr}$ are the eigenfunctions of the operator
$L$ with zero eigenvalues.
Let ${\bf P}_N$ be a  projection operator
on $E_N$ and ${\bf Q}_N={\bf I} - {\bf P}_N$.
The components of ${\bf P}_N$ are defined by
\begin{equation}
{\bf P}_{1, N} v= \sum_{j=1}^{N} \langle \tilde \omega, \tilde \omega_j^{+} \rangle \omega_j^{+} +   \sum_{j=1}^{N}\langle \tilde \omega, \tilde \omega_j^{-} \rangle \omega_j^{-},
\label{Pr2}
\end{equation}
\begin{equation}
{\bf P}_{2, N} v =(2\pi h)^{-1} \langle w, 1\rangle +  \sum_{j=1}^{N} \langle w, \tilde \theta_j ^+ \rangle \theta_j^+   +\sum_{j=1}^{N} \langle w, \tilde \theta_j ^- \rangle \theta_j^-,
\label{Pr2a}
\end{equation}
where  $v=(\tilde \omega, w)^{tr}$ and  $\tilde e_j^{\pm} =(\tilde \omega_i^{\pm}, \tilde \theta_i^{\pm})$ are eigenfunctions of the conjugate operator $L^*$ with zero eigenvalues $\lambda=0$
found above (see subsection  \ref{sec: 8.4}).

First we transform equations (\ref{eveq10})-(\ref{eveq11}) to a standard system with "fast" and "slow" modes.
 Let us introduce auxiliary functions $R_{\omega}(X)$, $R_{\psi}(X)$ and $R_{w}(X)$ by
$$
 R_{\omega}(X) =\sum_{j=1}^{N} X_j^+ \omega_j^+ + \sum_{j=1}^{N} X_j^- \omega_j^-,  \quad  R_{\psi}(X)=\sum_{j=0}^{N} X_j^ + \psi_j^+ +   \sum_{j=1}^{N} X_j^ -  \psi_j^-, 
$$
$$
 R_w(X)=X_0 +  \sum_{j=1}^{N} X_j^+ \theta_j^+    +  \sum_{j=1}^{N} X_j^- \theta_j^- ,
$$
and represent $\tilde \omega, \tilde \psi$ and $w$ by
\begin{equation}
\tilde \omega=\tilde  \gamma R_{\omega}(X) + \hat \omega,  \quad 
 \quad \tilde \psi= \gamma R_{\psi}(X) +\hat \psi,   
\label{om2}
\end{equation}
\begin{equation}
w= \gamma R_w(X) +\hat w,  
\label{w2}
\end{equation}
where
$
{\bf P}_N (\hat \omega, \hat w)^{tr}=0,
$
$X_i^{\pm}(t)$ and $\hat \omega, \hat w, \hat \psi$ are new unknown functions,
 $X=(X_0, X_1^+, ..., X_{N}^+, X_1^-,..., X_N^{-})^{tr}$.  

We  substitute relations (\ref{om2})-(\ref{w2}) in Eqs. (\ref{eveq10}) and (\ref{eveq11}).
 As a result,   one obtains  the system
\begin{equation}
\frac{dX_i^{\pm}}{dt}= \gamma^{-1}  (G_i^{\pm}( X) +  M_{i}^{\pm}( X) + f_i^{\pm} + F_i^{\pm}( X, \hat \omega,\hat w, \gamma)),
\label{X2}
\end{equation}
\begin{equation}
\hat \omega_t=\nu \Delta \hat \omega +  {\bf P}_{1,N} F(X, \hat \omega,\hat w, \gamma),
\label{hom2}
\end{equation}
\begin{equation}
\hat w_t= \Delta \hat w   - \{\hat \psi, U\} + {\bf P}_{2,N} G( X, \hat \omega,\hat w, \gamma),
\label{hw2}
\end{equation}
where
in Eqs.(\ref{hom2}) and (\ref{hw2}) 
\begin{equation}
F=\{\gamma R_{\psi}(X) + \hat \psi, \gamma R_{\omega}(X) + \hat \omega\},
\label{F2}
\end{equation}
\begin{equation}
G=\{\gamma R_{\psi}(X) + \hat \psi, \gamma R_{w}(X) + \gamma u_1 + \hat w\} 
+\gamma^2 \eta_1.
\label{G2}
\end{equation}
The functions $G_i^{\pm}(X)$ and $M_i^{\pm}(X)$ give main contributions in the right hand sides of Eqs.(\ref{X2}) and $F_i^{\pm}$ are corrections defined by 
\begin{equation}
F_i^{\pm}= \gamma^{-1} (F_i^{\pm, \omega} + F_i^{\pm, w}) , 
\label{Fi2}
\end{equation}
where
$$
F_i^{\pm, \omega}=\langle \{  \gamma R_{\psi}(X), \hat \omega \} + \{ \hat \psi, \gamma R_{\omega}(X)  + \hat \omega \}, \tilde \omega_i^{\pm} \rangle, 
$$
$$
F_i^{\pm, w}=\langle  \{  \gamma R_{\psi}(X), \hat w \}   + \{\hat \psi, \gamma R_{w}(X)  + \gamma u_1 + \hat w \}, \tilde \theta_i^{\pm} \rangle). 
$$

One has 
$$
G_i^{\pm}(X)=\langle\{R_{\psi}(X), R_{w}(X)\}, \tilde \theta_i^{\pm} \rangle +  \langle\{R_{\psi}(X), R_{\omega}(X)\}, \tilde \omega_i^{\pm} \rangle,    
$$
$$
M_i^{\pm}(X)=\langle\{R_{\psi}(X), u_1\}, \tilde \theta_i^{\pm} \rangle.
$$
These terms can be rewritten in a more explicit form as
\begin{equation}\label{Gpm}
  G_i^{+}( X) =\sum_{j,l=1}^N G_{ijl}^{+++} X_j^+ X_l^+  +  G_{ijl}^{--+} X_j^- X_l^-,  
\end{equation}
\begin{equation}\label{Gpm-}
    G_i^{-}( X) = \sum_{j,l=1}^N G_{ijl}^{+--} X_j^+ X_l^-
\end{equation}
and
\begin{equation}\label{Mpm}
  M_i^{+}( X) = \sum_{j=1}^N M_{ij}^{++} X_j^+
   + \sum_{j=1}^N M_{ij}^{+-} X_j^- , 
\end{equation}
\begin{equation}\label{Mpm-}
 M_i^{-}( X) =\sum_{j=1}^N M_{ij}^{-+} X_j^+
   + \sum_{j=1}^N M_{ij}^{--} X_j^-  .
\end{equation}
 Note that in Eqs. (\ref{Gpm}) - (\ref{Mpm-}) 
 all the other possible  terms vanish  since they are defined by integrals over $x$ of functions  odd in $x$.
The coefficients in (\ref{Gpm}),(\ref{Gpm-}), (\ref{Mpm}) and (\ref{Mpm-}) are defined by  
\begin{equation}
   M_{ij}^{\pm \pm}(u_1)=
\langle \{\psi_j^{\pm}, \tilde \theta_i^{\pm}  \},  u_1\rangle,
\label{MatrM}
\end{equation}
\begin{equation}
   G_{ijl}^{+ + +}=\langle \{\psi_j^{+}, \theta_l^{+}\}, \tilde \theta_i^{+} \rangle + O(\nu^{-1})
\label{G+}
\end{equation}
\begin{equation}
   G_{ijl}^{- - +}=\langle \{\psi_j^{-}, \theta_l^{-}\}, \tilde \theta_i^{+} \rangle + O(\nu^{-1})
\label{G-}
\end{equation}
\begin{equation}
   G_{ijl}^{ + --}=\langle \{\psi_j^{+}, \theta_l^{-}\} + \{\psi_l^{-}, \theta_j^{+}\}, \tilde \theta_i^{-} \rangle + O(\nu^{-1})
\label{matrG}
\end{equation}
for large $\nu$.   These estimates are obtained in \cite{Arch} by expressions for conjugate eigenfunctions.   One has
\begin{equation}
  f_{i}^{\pm}=
\langle \eta_1,  \tilde \theta_i^{\pm} \rangle.
\label{fipm}
\end{equation}

We consider Eqs.(\ref{X2}), (\ref{hom2}) and (\ref{hw2})  in the domain
\begin{equation}
   D_{\gamma, R, C_1, C_2} =\{(X, \hat w, \hat {\omega}): \ |X| <  R, \ ||\hat \omega||_{\alpha} < C_1\gamma^{2},
   ||\hat w||_{1 +\alpha}  < C_2\gamma^{2} \},
\label{Dom}
\end{equation}
where $\alpha > 3/4$.
Using results \cite{Arch}, one can prove the following assertion describing the normal form of dynamics in a small neighborhood of the bifurcation point.

\begin{lemma} \label{7.1}
{
Let $r  \in (0,1)$ and $|b-b_c| < \gamma^2$.
Assume $\gamma > 0$ is small enough:
$
\gamma < \gamma_0(N, \nu, R, r).
$
Then the local semiflow $S^t$, defined by equations  (\ref{X2}),(\ref{hom2}), and (\ref{hw2}) has a locally invariant
and locally attracting manifold ${\mathcal M}_{2N+1,\gamma}$. This manifold is  defined by
\begin{equation}
 \hat \omega= \hat \omega_0(X, \gamma),   \quad \hat w= 
   \hat w_0(X, \gamma),
\label{rW}
\end{equation}
where
$\hat \omega_0(X, \gamma)$, $\hat w_0(X,\gamma)$ are maps from the ball ${\mathcal B}^{2N+1}(R)=\{X: |X| <  R\}$ to $H_{\alpha}$ and $H_{1+\alpha}$ respectively,
bounded in $C^{1+r}$ -norm :
\begin{equation}
  |\hat \omega_0(X, \gamma)|_{C^{1+r}({\mathcal B}^{2N+1}(R))} < C_3\gamma^2, \quad |\hat w_0(X, \gamma)|_{C^{1+r}({\mathcal B}^{2N+1}(R))} < C_4 \gamma^2.
\label{rWe}
\end{equation}

The restriction of  the semiflow $S^t$ on ${\mathcal M}_{2N+1, \gamma}$  is defined by
\begin{equation}
  \frac{dX_i^{\pm}}{dt}=\gamma  (G_i^{\pm}( X) +  M_{i}^{\pm}( X) + f_i^{\pm}) 
    + \phi_i^{\pm}(X, \gamma)),
\label{maineq1}
\end{equation}
\begin{equation}
  \frac{dX_0}{dt}=0,
\label{maineq0}
\end{equation}
and the corrections $\phi_i^{\pm}(X, \gamma)=
 F_i^{\pm}( X, \hat \omega_0(X,\gamma),\hat w_0(X, \gamma), \gamma)
$ satisfy the estimates
\begin{equation}
  |\phi_i^{\pm}|, |D_X \phi_i^{\pm}|  < c_1 \gamma^s,   \quad  s>0.
\label{maineq4}
\end{equation}
}
\end{lemma}

Our aim in coming sections is to prove that the normal forms (\ref{maineq1}) exhibit very complex dynamics. To be precise this assertion let's us formulate the definition.

{\bf Definition} {\em  Consider a  family of the semiflows $S^t(P)$, where $P$ is a parameter, 
  defined on Banach space $H$. If this family $\epsilon$- realizes (in the sense of subsection \ref{RVF}) all smooth finite dimensional systems with arbitrarily prescribed accuracy $\epsilon >0$ then we say  that family is is maximally
dynamically complex.}  

The meaning of this definition is that maximally dynamically complex systems can generate all kinds of structurally stable dynamics, in particular, all hyperbolic dynamics. Such dynamics may be chaotic
(as examples, we can take Anosov's flows,  homoclinic chaos and others \cite{Holmes,Katok}).  To formulate an important property of maximally dynamically complex systems, 
let us denote by ${ B}^n(R)$ be the  ball $\{q: |q|\le R\}$ in ${\bf R}^n$ of the radius $R>0$ centered at $0$, where $q=(q_1, q_2, ..., q_n)$ and $ |q|^2=q_1^2 + ...
+ q^2_n$ and 
consider a system of differential equations on the  ball ${ B}^n(R)$:
\begin{equation}
 \frac{dq}{dt}=Q(q),
\label{ordeq}
\end{equation}
 where
\begin{equation}
   Q \in C^1({ B}^n), \quad \sup_{q \in { B}^n(R)}|\nabla Q(q)| < 1.
\label{cond1}
\end{equation}

 Suppose the vector field $Q$ is directed strictly
inward to the  ball ${ B}^n(R)$ at its boundary $\partial { B}^n(R)=\{q: |q|=R \}$:
\begin{equation}
   Q(q) \cdot q < 0 \quad for \ q \in \partial { B}^n(R).
\label{inward}
\end{equation}
Then Eq. (\ref{ordeq}) defines a finite dimensional  global semiflow on ${ B}^n(R)$.
\vspace{0.2cm}
Now we can formulate

{\bf Proposition}. {\em If a family of the semiflows is maximally dynamically complex, 
then these semiflows  enjoy the following property. 
For each integer $n$, each $\epsilon > 0$  and each vector field
$Q$ satisfying (\ref{cond1}) and (\ref{inward}) and  having a hyperbolic dynamics $G_{\Gamma}^t$ on a compact invariant hyperbolic set $\Gamma$, there exists a value of the parameter
$\mathcal P$ such that
the corresponding system   (\ref{gks}) defines a  semiflow $S^t_{\mathcal P}$, which also has a hyperbolic dynamics $G_{\Gamma'}^t$  on a  hyperbolic set $\Gamma'$, which is homeomorphic to  $\Gamma$. The dynamics $G_{\Gamma^{'}}^t$ and $G_{\Gamma}^t$
are orbitally topologically equivalent}.

For definitions of hyperbolic sets, dynamics  and orbital topological equivalence  see, for example,
\cite{Holmes,Katok} among others. 
  
The proof of this claim  (see \cite{Vakbook}) uses the theorem on persistence of hyperbolic sets \cite{Katok}. According to this theorem, we can choose a sufficiently small  $\epsilon(Q, \Gamma) >0$ such that if estimate (\ref{estRVF}) holds
then system (\ref{reddynam}) has a compact invariant hyperbolic set $\Gamma'$ homeomorphic to  $\Gamma$ and, moreover, the global semiflows defined by systems  (\ref{ordeq}) and (\ref{reddynam}) generate orbitally topologically equivalent dynamics on invariant sets  $\Gamma$ and $\Gamma^{'}$, respectively.  

The theory of maximally dynamical complex systems is based on the RVF method (on the RVF  method see Appendix, subsection \ref{RVF})
and it is  developed for neural networks and reaction diffusion systems (see \cite{Vakbook}). In this paper, we first state a hydrodynamical example.

In the next section we apply the RVF method 
for quadratic systems (\ref{maineq1}). As a parameter $P$  we use the function $u_1(x,y)$, i.e., a small
two dimensional spatial inhomogeneity. The matrix $M^{\pm}$ in the right hand side of (\ref{maineq1})  is a linear functional of $u_1$.
One can show (see \cite{Arch}) that, by adjusting $u_1$, we can obtain any matrices $M^{\pm}$. This assertions seems natural  since
the matrices $M^{\pm}$  contain $N^2$ entries, thus we should satisfy $N^2$ restrictions by an infinite set of unknowns, which are the Fourier coefficients of  
$u_1$. 

So, the matrix $M^{\pm}$  in normal forms (\ref{maineq1}) can take any values, however, the main difficulty is that the quadratic terms in the normal forms  (\ref{maineq1}) are not arbitrary and subject some restrictions.    
Our plan to overcome this difficulty is as follows. We use the key auxiliary assertion, Lemma \ref{quadr2}, which means that any given quadratic system can be realized by a normal form (\ref{maineq1}) of a sufficiently large dimension $N$.

\section{Reductions of quadratic systems}
\label{sec7}

Our next step is to study a general class of quadratic systems, which includes (\ref{maineq1}) as a particular case.
We consider the following systems
\begin{equation}
   \frac{dX}{dt}=   { K}(X)    + { M} X + g,
\label{gks}
\end{equation}
where  $X=(X_1, ..., X_N), \ { K}=(K_1, ...,K_N),  { g}=(g_1, ..., g_N) \in {\bf R}^N$, $K(X)$ is a quadratic map
defined by
$$
{K}_i(X)= \sum_{j=1}^N \sum_{l=1}^N K_{ijl} X_j X_l,
$$
and $MX$ is a linear operator
$
({ M}X)_i= \sum_{j=1}^N M_{ij} X_j.
$
System (\ref{gks}) defines a local semiflow $S^t(g,{ M})$ in the ball ${\mathcal B}^N(R_0) \subset {\bf R}^N$ of the radius $R_0$
centered at $0$. We shall consider the vector $g$ and the matrix $ M$ as parameters of this semiflow whereas
the entries $K_{ijl}$ will be fixed.

Let us formulate an assumption on entries $K_{ijl}$.
We present  $X$  as $X=(Y, Z)$, where
$$
Y_l=X_{i_l},  \quad l \in I_p, \quad Z_l=X_{j_l}, \quad l \in J_p,
$$
where  $I_p=\{i_1,..., i_p\}$ and $J_p=\{j_1, ..., j_{N-p} \}$ are disjoint subsets of $\{1,..., N\}$ such that
$
   I_p \cup J_p=\{1,..., N\}.
$
Then  system (\ref{gks}) can be rewritten as follows:
 \begin{equation}
   \frac{dY}{dt}=   { K}^{(1)} (Y) + { K}^{(2)} (Y, Z) + { K}^{(3)}(Z) + { R} Y + { P} Z
 +  f,
\label{gks2}
\end{equation}
 \begin{equation}
   \frac{dZ}{dt}=  \tilde { K}^{(1)} (Y) + \tilde{ K}^{(2)} (Y, Z) + \tilde { K}^{(3)} (Z)  +  \tilde { R} Y +  \tilde { P} Z
 +  \tilde f,
\label{gks3}
\end{equation}
where
\begin{equation}
   { K}^{(1)}_i(Y)=  \sum_{j \in I_p} \sum_{l \in I_p}  K_{ijl}^{(1)} Y_{j} Y_{l}, \quad \tilde { K}_k^{(1)}(Y)= \sum_{j\in I_p} \sum_{l \in I_p}  \tilde K_{kjl}^{(1)} Y_{j} Y_{l},
\label{KY1}
\end{equation}
\begin{equation}
   { K}^{(2)}_i(Y,Z)=  \sum_{j \in I_p} \sum_{l \in J_p}  K_{ijl}^{(2)} Y_{j} Z_{l}, \quad \tilde { K}_k^{(2)}(Y,Z)= \sum_{j \in I_p} \sum_{l \in J_p}  \tilde  K_{kjl}^{(2)} Y_{j} Z_{l},
\label{KY2}
\end{equation}
\begin{equation}
   { K}^{(3)}_i(Z)=  \sum_{j \in J_p} \sum_{l \in J_p}  K_{ijl}^{(3)} Z_{j} Z_{l}, \quad \tilde { K}_k^{(3)}(Z)=\sum_{j \in J_p} \sum_{l \in J_p}  \tilde  K_{kjl}^{(3)} Z_{j} Z_{l}
\label{KY3}
\end{equation}
and  where $i \in \{1,..., p\},   \    k \in \{1,..., N-p\}$.  Linear terms have the form
 \begin{equation}
 (\tilde { R} Y)_i= \sum_{j \in I_p} \tilde R_{ij} Y_j,  \quad (\tilde { P} Z)_i=\sum_{j \in J_p} \tilde P_{ij} {Z_j},
 \quad  ({ P} Z)_i= \sum_{j \in J_p} P_{ij} Z_j,
\label{gks3c}
\end{equation}
and $f=(f_{1}, ..., f_{p}), \ \tilde f=(\tilde f_1, ..., \tilde f_{N-p})$.
We denote by  $S^t({\mathcal P})$ the local semiflow defined by
(\ref{gks2}) and (\ref{gks3}). Here ${\mathcal P}$ is a semiflow parameter, $ {\mathcal P}=\{f, \tilde f, {P}, \tilde { P},
{R}, \tilde { R} \}$.
\vspace{0.2cm}

{\bf $p$-Decomposition Condition} \label{ass40}
{\em
Suppose  entries $K_{ijl}$ satisfy  the following condition.
For some $p$ there exists a decomposition $Z=(X, Y)$ such that
   for all  numbers $b_{jl}$, where $j,l \in I_p$, 
 the  linear system  
\begin{equation}
	     \sum_{i \in J_p} \tilde K_{ijl}^{(1)}  u_i= b_{jl}, \quad l, j \in I_p
\label{barK}
\end{equation}
has a solution $u$.
}
\vspace{0.2cm}

Clearly, for $N > p^2 + p$ and  generic matrices $K$ this condition is valid.

Let us formulate  some conditions to  the matrices ${ R}, \tilde { R}, {P}$ and $\tilde { P}$.
Let $\xi >0$ be  a  parameter.  We suppose that
\begin{equation}
  \tilde P_{ij}=-\xi^{-1} \delta_{ij},    \quad  i=1,..., N-p, \ j=1, ...,
\label{gks4c}
\end{equation}
where $\delta_{ij}$ is the Kronecker symbol,
\begin{equation}
  \tilde R_{ij}=0, \quad  \tilde f_i=0, \quad i=1,..., N-p, \ j=1,...,p,
\label{gks4d}
\end{equation}
\begin{equation}
   P_{ij}=\xi^{-1} T_{ij}, \quad |T_{ij}| < C_0, \quad i=1,..., p, \ j=1,..., N-p,
\label{gks5c}
\end{equation}
\begin{equation}
 |R_{ij}| < C, \quad i=1,..., p, \ j =1,..., p,
\label{gks5d}
\end{equation}
Let us fix $p$ and consider the numbers $N, \xi$, coefficients $ T_{ij}, R_{ij}$ and $f_i$ as a parameter $P$. 

\begin{lemma} \label{quadr1} {Assume  (\ref{gks4c}),  (\ref{gks4d}), (\ref{gks5c}) and (\ref{gks5d})  hold.
For sufficiently small positive $\xi < \xi_0(R_0, r, f)$ the local semiflow $S^t(P)$ defined by  system (\ref{gks2}), (\ref{gks3}) 
has a locally invariant and locally attracting manifold ${\mathcal M}_{P}$.  This manifold is  defined by equations
\begin{equation}
  Z= \xi (\tilde { K}^{(1)}(Y)  +  W(Y,\xi)), \quad Y \in { B}^p(R_0)
\label{gks6c}
\end{equation}
where $W$ is a $C^1$ smooth map from the  ball ${B}^p(R_0)$ to ${\bf R}^{N-p}$ such that for some $C_1,s>0$
\begin{equation}
|W(\cdot, \xi)|_{C^1({ B}^p(R_0))}  < C_1\xi^s.
\label{gks7c}
\end{equation}}
\end{lemma}

{\bf Proof}. 
 Let us introduce a new variable $w$ by
\begin{equation}
  Z= \xi (\tilde { K}^{(1)}(Y)  +  w).
\label{gks8c}
\end{equation}
and the rescaled time by $t=\xi\tau$.
Then for $Y, w$ one obtains the following system
\begin{equation}
   \frac{dY}{d\tau}=   \xi G(Y, w, \xi),
\label{gk1}
\end{equation}
 \begin{equation}
   \frac{dw}{d\tau}=  \xi F(Y, w, \xi)
    -  w,
\label{gk2}
\end{equation}
where
$$
    G(Y, w, \xi)={ K}^{(1)} (Y) + \xi {K}^{(2)} (Y, \tilde { K}^{(1)}(Y)  +  w) +
$$
$$
    +
  \xi^2 { K}^{(3)}(\tilde { K}^{(1)}(Y)  +  w) + { R} Y + { T}(\tilde { K}^{(1)}(Y)  +  w)
 +  f,
$$
$$
    F(Y, w, \xi)=   \tilde{ K}^{(2)} (Y, \tilde { K}^{(1)}(Y)  +  w)+ $$
 $$   +
 \xi  \tilde { K}^{(3)} (\tilde { K}^{(1)}(Y)  +  w) +  h(Y, w, \xi),
$$
\begin{equation}
    h(Y, w, \xi)= -D_Y \tilde { K}^{(1)}(Y)  G(Y, w, \xi).
\label{gk4}
\end{equation}

 Equations (\ref{gk1}), (\ref{gk2}) form  a typical system involving slow ($Y$) and fast ($w$) variables. 
Existence of a locally invariant manifold for  this system can be shown by the well-known results 
(see \cite{He}). The remaining part of the proof is standard (see \cite{Arch}) and is omitted $\square$.

The semiflow $S^t$ restricted to ${\mathcal M}$ is defined by the equations
\begin{equation}
\frac{dY}{dt}=   F(Y,  \xi),
\label{inert}
\end{equation}
where
$$
    F(Y,  \xi)={ K}^{(1)} (Y) + \xi { K}^{(2)} (Y, \tilde { K}^{(1)}(Y)  +  W(Y, \xi)) +
$$
$$
    +
   \xi^2 { K}^{(3)}(\tilde { K}^{(1)}(Y)  +  W(Y,\xi)) + { R} Y + { T}\tilde { K}^{(1)}(Y)  + W(Y, \xi))
 +  f.
$$
The estimates for $W$ show that $F$ can be presented as
\begin{equation}
F(Y,  \xi)={ K}^{(1)} (Y)  + { R} Y + { T}\tilde { K}^{(1)}(Y) + f + \phi_{\xi}(Y)
\label{inert2}
\end{equation}
where a small correction $\phi_{\xi}$ satisfies
\begin{equation}
|\phi_{\xi}|_{C^1({ B}^p(R_0))} < c_0\xi^{1/2}.
\label{inert2c}
\end{equation}
In (\ref{inert2})  $ R$ and $f$ are free parameters. The quadratic form  ${ D}(Y)={ K}^{(1)} + { T}\tilde { K}^{(1)}$
 can be also considered as  a free parameter according to the $p$- Decomposition Condition.
Therefore, we have proved the following assertion.

\begin{lemma} \label{quadr2} { Let
\begin{equation} \label{QField}
F(Y)= { D}(Y) + { R} Y + f
\end{equation}
be a quadratic vector field on ${\mathcal B}^p(R_0)$, where
$$
{D}_i(Y)=\sum_{j=1}^p \sum_{l=1}^p  D_{ijl} Y_j Y_l, \quad ({ R} Y)_i=\sum_{j=1}^p R_{ij} Y_j.
$$
Consider system  (\ref{gks2}), (\ref{gks3}). Let the $p$- Decomposition Condition hold.
Then for any $\epsilon >0$ the field $F$ can be $\epsilon$ - realized by the semiflow $S^t({\mathcal P})$ defined  by
system (\ref{gks2}), (\ref{gks3}), where parameters $\mathcal P$ are the dimension $N$, the matrices ${ P}$, $ R$, $ \tilde  P$,
$ \tilde  R$ and the vectors $f, \tilde f$.
}
\end{lemma}

This lemma immediately follows from the main result on quadratic systems (\ref{gks}). Note that the class of
systems (\ref{QField}) includes the Lorenz model as a particular case.

\begin{theorem} \label{maint}{ 
Consider the family of semiflows defined by  systems  (\ref{gks}),   where the triple $\{N, M, g \}$ serves as a parameter ${\mathcal P}$,  
and for each $N$ the coefficients $K_{ijl}$ with $i, j, l \in \{1,..., N\}$ satisfy $p$-decomposition condition for an integer $p$ such that $N/2 < p^2 + p \le N$.  Then that family is maximally dynamically complex 
}.
\end{theorem}

  {\bf Proof}.
According to results \cite{Stud} for any $\epsilon_1 >0$ we can construct $\epsilon_1$-realization of the field $Q$ by semiflows defined by (\ref{QField}). Moreover,  due to the previous lemma, 
for any $\epsilon_2 >0$ we can find   $\epsilon_2$ -realization of any system
(\ref{QField}) by semiflows defined by (\ref{gks}).  If $\epsilon_k>0$ are small enough, these two realizations give us $\epsilon$-realization of $Q$.  The corresponding system (\ref{gks}) has the hyperbolic set $\Gamma'$.  This completes the proof.

The last result show that quadratic systems (\ref{gks}), that arise as a result of our bifurcations, can exhibit all chaotic hyperbolic dynamics, for example, Anosov \'s flows or axiom A  Smale dynamics.
To prove that such hyperbolic chaotic dynamics can be generated by the original Marangoni fluid dynamics (defined by IBVP (\ref{OBEstream2})-(\ref{Initdata})) we need additional technical assertions.  They can be obtained following
\cite{Arch}.

\section{Concluding remarks}
\label{sec: 12}

Bifurcations of fluid dynamics leading to periodic spatial patterns (for example, B\'enard cells) or time periodic regimes are well studied.
In 1971, Ruelle and Takens \cite{NRT} pioneered the hypothesis that there are more complicated bifurcations possible, which describe a transition from a simple rest point attractor 
to strange attractors (which can describe a turbulence). However, until now there exists no completely analytical proof of the existence of such bifurcations.    
Note that turbulence exhibits not only complicated time dynamics, but complex spatial patterns are also observed.

This paper states a proof of existence of  bifurcations, which can produce turbulence and complicated  patterns.  
It is shown that, at the bifurcation point, the dynamics  can be described by systems of differential equations with quadratic nonlinearities. Such systems can have
chaotic dynamics, specifically, they can generate all finite dimensional hyperbolic dynamics.

 Physically the complicated spatio-temporal patterns are generated by 
diffusion, convection, and the capillary effect.  The fact that the Marangoni effect can induce an interfacial turbulence has been long known
from experiments and numerical simulations (see, for example, \cite{turb, turb2}). In our model, the physical mechanism of this phenomenon
is an interaction of slow modes which determine the dynamics and the spatial inhomogeneities in the system.  

We think that this mechanism is also applicable for climate models. Indeed, climate systems always include slow and fast components. Therefore, in the climate models
there are  internal interactions between 
the slow segments of the climate system and  
 the fast weather components. The results of this paper 
show that these interactions can lead to different variants of complex dynamics.  We have a new physical mechanism for the generation of complicated dynamics. Namely, it is shown that in spatially extended systems with many slow variables  different small inhomogeneities can lead to sharply different dynamics, which may be chaotic. Mathematically it can be shown by the method of realization of vector fields, however, the physical idea behind this method is  transparent. In fact, in spatially extended systems the number of slow modes usually is much smaller than the number of the fast ones.  Some space inhomogeneities define an interaction between the  fast and slow modes. We have thus a  number of parameters which affect the dynamics of the slow variables and which can be used to control those dynamics.

\section*{Acknowledgments}

We gratefully acknowledge support from the RFBR under the Grant \#16-31-60070 mol\_a\_ dk and from the Government of the Russian Federation through mega-grant 074-U01. This research was also supported by the German Federal Ministry of Education and Research (BMBF) through the "Green Talents" Program. IS acknowledges the kind hospitality of the Isaac Newton Institute for Mathematical Sciences (Cambridge, UK) and of the Mathematics for the Fluid Earth 2013 Program. In preparing this text, we have benefited from discussions with Prof. Valerio Lucarini (University of Reading).

\section{Appendix} 

\underline{\bf{~~~~~~~~~~~~~~~~~~~~~~~~~~~~~~~~~~~~~~~~~~~~~~~~~~~~~~~~~~~~~~~~~~~~~~~~~~~~~~~~~~~~~~~~~~~~~~~~~~}}

\subsection{ Proof of Theorem \ref{6.2} }

{\bf  Proof}.  First
we use  Lemma \ref{6.1} to obtain a nonlinear equation for the eigenvalues $\lambda(k)$ of the boundary value
problem (\ref{BVK1})-(\ref{BVK3}). As a result,    one has
\begin{equation}
k^2 \nu \lambda^{-1} D^{-1} \beta_k \int_0^h   \Phi_k(y, \lambda) \psi_k(y,\lambda) \rho_{\bar k}(y) U_y(y, \nu)   dy=b^{-1} \beta_k(\lambda).
\label{kwc2}
\end{equation}

Note that the operator $L$ is not self-adjoint. Therefore, complex eigenvalues $\lambda$ may appear, i.e., complex roots of (\ref{kwc2}).
Moreover, let us note that, according to (\ref{psiF2}), if $\beta_k( \lambda)=0$, then Eq. (\ref{kwc2}) is satisfied.
In this case  (\ref{BVK1}) entails that
\begin{equation} \label {BVK10}
\frac{d^2\omega_{k}(y)}{dy^2} - k_{\nu}^2 \omega_k=0,   \quad  \omega_k(h)=0, \quad \omega_k(0)=0,
\end{equation}
therefore $k_{\nu}^2= -(n \pi/h)^2$, where $n$ is an integer. This gives
$\lambda=-\nu((n \pi/h)^2+ k^2)< -1/2$. These eigenvalues $\lambda$ correspond to trivial solutions of the eigenfunction problem with  $\lambda \notin {\bf C}_{1/2}$.  
Therefore,   without loss of generality we can set
$\beta_k(\lambda)=1$ in Eq.(\ref{kwc2}).

The plan of the remaining part of the proof is as follows.
We consider the two  cases:
$
( {\bf I}) \hspace{0.1cm} |\lambda| < \nu^{3/4}
$
and
$
 ({\bf II}) \hspace{0.1cm} |\lambda| > \nu^{3/4}.
$
In the  first case we can simplify equation (\ref{kwc2}), while 
in the second case a rough estimate shows that
 Eq. (\ref{kwc2})  has no solutions.

Let us start with the case {\bf I}. To simplify our statement, 
we  first consider a formal limit of Eq.(\ref{kwc2}) as $\nu \to +\infty$. Using (\ref{psiFa1}), (\ref{tildepsi}) and (\ref{rhoest1}) one obtains that this limit has the form 
\begin{equation}
(2\bar D)^{-1} \int_0^{+\infty}   y \bar U_y(y) \exp(-(k +\bar k) y)dy=\bar k/k, \quad  \bar D=Db^{-1},
\label{kwc3}
\end{equation}
where  $\bar U$ is the  limit  $\bar U=\lim  U_{K_N}(y, \nu)$ as $\nu \to +\infty$ (it will be shown below that this limit exists).
We  set $\bar U=V(y, d)$, where $V$ is a function  of a special form 
that depends on some parameters $d=(d_1,..., d_M)$,  where $M=k_N$.
Consider $C^{\infty}$ - mollifiers
$\delta_{\epsilon}(y)$ such that $\delta_{\epsilon} \ge 0$,  the support  $supp \ \delta_{\epsilon}(y)$ is
$(-\epsilon, \epsilon)$ and
\begin{equation}
\int_{-\epsilon}^{\epsilon}\delta_{\epsilon}(y)dy=1,
\label{Inteps}
\end{equation}
\begin{equation}
\sup |D_y^k\delta_{\epsilon}(y)| < c_k \epsilon^{-(k+1)}, \quad  k=0,1,2.
\label{Inteps1}
\end{equation}

Let us define  the function $V(y,d)$ on $[0, \infty)$ by 
\begin{equation}
 V_y(y, d)=2y^{-1}(\delta_{\kappa}(y-z_0) + \mu y P_M(y, d)),   
\label{Uy}
\end{equation}
where $\chi(z)$ is the step function such that
$
\chi(z)=1$ for  $z >0$ and  
$\chi(z)=0$ for $ z < 0$,
$P_M$ is a polynomial in $y$ of the degree $M+1$ with coefficients depending on $(d_1, d_2, ..., d_M)$,
and
\begin{equation}
  \mu = \kappa^{2/3}, \quad z_0=5\kappa,     
\label{z0}
\end{equation}
where $\kappa$ is a small parameter independent of $\nu$ and $\gamma$. 
Assume that coefficients of the polynomial $P_M$ and $d_j$ are  bounded:
\begin{equation}
P_M(y, d)=\sum_{j=0}^{M} b_j(d) y^j, \quad  |b_j(d)| < C,   \quad  |d_j| < 1/2.
\label{PN}
\end{equation}

To investigate (\ref{kwc3})
 it is useful to  introduce the variable
\begin{equation}
p=k +\bar k=k +(k^2 +\lambda)^{1/2}.
\label{pk}
\end{equation}
 Then  Eq. (\ref{kwc3}) can be rewritten as
 \begin{equation}
  \bar D  \frac{p}{k}=1 + \bar D  + S(p, k, d),
\label{ME}
 \end{equation}
 where
\begin{equation}
    S(p, k, d)=\mu G(p, d)  + g_{\kappa}(p) ,
\label{kwe70}
 \end{equation}
 \begin{equation}
  g_{\kappa}(p)=  -1  + \int_0^{\infty} \delta_{\kappa}(y-z_0) \exp(-py) dy,
\label{kweg}
 \end{equation}
 and
 \begin{equation}
    G(p, d)=\int_{0}^{\infty} y P_M(y, d) \exp(-py) dy.
\label{kwe71}
 \end{equation}

We suppose that $\bar k >0$ thus
$Re \ p > k$.
Therefore, we  can  investigate (\ref{ME}) in the domain
\begin{equation}
{\bf C}_{1/2, k}=\{p \in {\bf C}: \ p=\sqrt{k^2 +\lambda}+k,  \  Re \ \lambda > -1/2,  \  Re \ p > k \}.
\label{domp}
\end{equation}
Note that
\begin{equation} \label{lamviap}
Re \ p= k + \sqrt{k^2 + Re \lambda  + (Im \ p)^2}, 
\end{equation}
 this shows that in ${\bf C}_{1/2, k}$ we have $Re \ p > 2k -1/2$.
We also assume
\begin{equation} \label{Dk}
|\bar D - 1|  \le  \kappa.
\end{equation}
Taking into account this restriction to $\bar D$, we define the interval $J=(b_1, b_2)$ and $b_c$ by  
\begin{equation} \label{Dkb}
b_1=\bar D^{-1}(1-\kappa),   \quad  b_2=\bar D^{-1}(1 +\kappa),  \quad b_c=\bar D^{-1}. 
\end{equation}

We can choose a polynomial $P_M$ such that 
\begin{equation}  \label{Gpd}
G(p,d)=p^{-2} (-1)^{M+1} \prod_{j=1}^M (\frac{1}{p} -  \frac{1}{2j + d_j}).
\end{equation}

Let us formulate an  auxiliary assertion.

\begin{lemma} \label{6.3} {
One has
\begin{equation}
 Re \ g_{\kappa}(p) \le - \min \{2 \kappa Re \ p, \quad 1/2   \}.
\label{g0}
\end{equation}  
}
\end{lemma}

{\bf Proof}. Estimate (\ref{g0})  follows from   (\ref{z0}) and (\ref{kweg}).
Indeed,  due to (\ref{Inteps}) we have
$$
Re \ g_{\kappa}(p) = \ \int_0^h \delta_{\kappa}(y-z_0) (Re \ \exp(-py)-1)dy\le $$
$$
\le \int_0^h \delta_{\kappa}(y-z_0) (\exp(-Re \ p \ y)- 1) dy
$$
that according to (\ref{z0})  gives 
\begin{equation} \label{Jps}
Re \ g_{\kappa}(p)  <  \exp( - 4 \kappa Re \ p )-1.
\end{equation}
Consider the function $J(x)=\exp(-x) -1$.  By the Taylor series we obtain that
$J(x) \le -x + x^2/2$ for $x \ge 0$.  This shows that $J(x) \le -1/2$ for $x >1$ and
$J(x) \le -x/2$ for $x \in [0,1)$.  These inequalities and (\ref{Jps}) imply 
(\ref{g0}) $\square$.

Let us show that  in the case $ |p| > \kappa^{-3/4}$,  under assumption (\ref{Dk}), Eq. (\ref{ME}) has no solutions with $Re \lambda >- c_0  \kappa$.

\begin{lemma} \label{plarge}
{If $|p| > \kappa^{-3/4}$ then for sufficiently small $\kappa$ solutions of (\ref{ME}) satisfy
\begin{equation}  \label{Repest}
Re \ p  < 2k  -c_4 k^2 \kappa^{1/4}, \quad c_4 >0.
\end{equation}
For the corresponding $\lambda_k(p)$ one has
\begin{equation} \label{lamest}
Re \ \lambda_k < -c_5 k^2 \kappa^{1/4}, \quad c_5 >0.
\end{equation}
} 
\end{lemma}

{\bf Proof}. Relation (\ref{lamviap}) shows  that for $p \in {\bf C}_{1/2, k}$ one has $Re \ p > Im \ p$,   thus $|p| > \kappa^{-3/4}$ entails  $4Re \ p >  \kappa^{-3/4}$.  
Then estimate (\ref{g0})  implies
$
 Re \ g_{\kappa}(p)    < -c_{2}\kappa^{1/4}.
$
Moreover,  
$
 \mu Re  \  G(p, d)    < -c_{3} \mu.
$
These  estimates and condition  (\ref{Dk})  entail   that  for small $\kappa$ one has
$Re \ S(p, k, d) < -c_4 \kappa^{1/4}$ and, 
therefore,
(\ref{Repest}) holds.  By (\ref{lamviap}) this gives us (\ref{lamest}).
$\square$

Consider the case $|p| < \kappa^{-3/4}$.
Let us introduce a new unknown $\tilde p$ by
$
(2+\tilde p)k= p.
$
Then equation (\ref{ME}) can be  rewritten  as
\begin{equation}
\tilde p  =H(\tilde p, k, \bar D),
\label{ME1}
 \end{equation}
where
\begin{equation}
H(\tilde p, k, \bar D)= \bar D^{-1}( g_{\kappa}((2+\tilde p)k)  + \mu G( (2+ \tilde p)k, d))  + \bar D^{-1} -1.
\quad \label{Hpk}
 \end{equation}
Let us prove an  estimate of solutions to (\ref{ME1}).

 \begin{lemma} \label{6.4} { In the domain ${\mathcal D}_{\kappa, k}=\{ p: p \in {\bf C}_{1/2, k}, \ |p| < c_1\kappa^{-3/4} \}$ solutions of (\ref{ME1})
 satisfy
 \begin{equation}
 |\tilde p | < C_{1}\kappa^{1/4},  \quad C_1 >0.
 \label{tp}
 \end{equation}
 }
 \end{lemma}

{\bf Proof}.
 To prove this  lemma, we note that if $p \in {\mathcal D}_{\kappa, k}$ then by
(\ref{kwe70}),(\ref{kweg}),  and  (\ref{kwe71})
one has
\begin{equation}
| g_{\kappa} (p)| < C_2\kappa^{1/4}, \quad |G(p, d)| < C_3.
\label{g0G}
\end{equation}
 Therefore, using (\ref{Dk}) and (\ref{z0}) one obtains
\begin{equation} \label{Hunif}
|H(\tilde p, k)| < C_5 (\mu + \kappa^{1/4}).
\end{equation}
Now (\ref{ME1}) shows that $\tilde p$  satisfies (\ref{tp}) $\square$.

Let us consider  equation (\ref{ME1}).  Using the last lemma we note that, to resolve this equation,  
we can apply  a  perturbation theory.
Relations  (\ref{kwe70}), (\ref{kweg}), and (\ref{Gpd}) show that  in the  domain ${\mathcal D}_{\kappa, k}$ 
one has
$$
|\frac{\partial H(\tilde p, k)}{\partial \tilde p}|  < C_6  \mu
$$
for $p$ satisfying  (\ref{tp}). 
Now Lemma \ref{6.4} and
 the implicit function theorem  entail  that for sufficiently small $\kappa$ all roots $\tilde p$ of  eq.  (\ref{ME1}) lie in ${\mathcal D}_{\kappa, k}$ and can be found by 
contracting mappings.  For each fixed $k$  the solution $\tilde p_k$ of eq.  (\ref{ME1})  is unique in ${\mathcal D}_{\kappa, k}$.

Let us set 
\begin{equation}
\tilde p_k^*= -  \kappa, \quad  k \in \{ 1, ..., M\}, \ and \  k \ne k_j,  
\end{equation}
and
\begin{equation}
\tilde p_k^*=0, \quad k = k_j,  \quad j=1,..., N. 
\end{equation}

\begin{lemma}   \label{choiced}
{Let us define $b_1$, $b_2$ and $b_c$  by (\ref{Dkb}).  Then for sufficiently small $\kappa$ we can choose  $d_j$, $j=1,..., M$, such that  $|d_j| < 1/2$ and:

{\bf A}) for each $k \in \{1, ..., M \}$ and $b=b_c$ eq. (\ref{ME1}) has a unique solution $\tilde p=\tilde p_k^*$;

{\bf B})  for $k=k_j$, $j=1, ..., N$  one  has $\tilde p_k(b_c)=0$ and roots $\tilde p_k(b)$ change their signs at $b=b_c$ as $b$  goes through the interval $(b_1, b_2)$;

{\bf C})   for $k > M$   and $b \in (b_1, b_2)$ solutions $\tilde p_k$ of eq. (\ref{ME1}) satisfy  
$$
 \tilde p_k  < -c \kappa^s, \quad s >0.
$$}
\end{lemma}
   
{\bf Proof}.
Let us fix $k \in \{1,..., M\}$ and consider $p \in {\mathcal D}_{\kappa, k}$, i.e., $p$ close to $2k$. Then  for $\tilde p$ satisfying (\ref{tp}) from (\ref{Gpd}) we obtain the following asymptotic for $\tilde G(\tilde p, d)=
G((2+\tilde p)k, d)$: 
$$
\tilde G(\tilde p,d)= (-1)^{M+1} (\bar a_k + \tilde a_k(\tilde p ,d)) (d_k - k \tilde p),
$$
where
$$
\bar a_k = (4k^2)^{-2} \prod_{j=1, j \ne k}^{M} (\frac{1}{2k} -  \frac{1}{2j}),  
$$
and $\tilde a_k(\tilde p, d)$ is an analytic function such that $|\tilde a_k| =O(|\tilde p| + |d|)$ for small $\tilde p, d$. Therefore,   
eq. (\ref{ME1}) can be transformed to the form
\begin{equation}\label{ME1d}
\tilde p= R_k(\tilde p, d, \kappa)  +    \mu (-1)^{M+1} \bar a_k d_k,
\end{equation}
where $R_k$ is an analytic function of $\tilde p$  and $d$ for   small $|\tilde p|, |d|$  such that 
$$
 \sup_{\tilde p_k, d: |d| < \mu^{1/2}, \tilde p_k \in  {\mathcal D}_{\kappa,k}} (|R_k| + |grad_{\tilde p} R_k|) < c_1 \kappa,    
$$
for some $c_1 >0$. 
Therefore, for small $\kappa$  we can apply the Implicit Function Theorem to find $d_j$, $j=1,..., M$ such that the root $\tilde p_k$ of (\ref{ME1d}) satisfies condition  ({\bf A}).
The assertion {\bf B}) also follows from equation (\ref{ME1d}). 

 To prove assertion ({\bf C}),  let us consider an estimate of $Re \ \tilde p$ for  $k > k_N$.  We observe
that for $k > k_N$ and  $\tilde p$ satisfying (\ref{tp}) we have $\mu Re \ G < -c_2 \kappa^{2/3}$,  $Re \  g_{\kappa}(p) < -c \kappa$,      
and thus using (\ref{Dk}) and eq. (\ref{ME1}) we obtain $Re \ \tilde p <  - c_2 k \kappa^{2/3}$.  
$\square$.

\vspace{0.1cm}
Finally, we have obtained needed estimates of solutions (\ref{kwc2}) for the case {\bf I}, $\beta_k=1$  and  $\nu=+ \infty$.
To finish our investigation of Eq.(\ref{kwc2})  for the case of large $\nu$,  we compare Eqs. (\ref{kwc2}) and its formal limit (\ref{kwc3}).  
We  assume that the function $U=V(y, d)$ is defined as above, by  (\ref{Uy}).

We observe that  for $\beta_k=1$ Eq. (\ref{kwc2}) can be rewritten as
\begin{equation}
\int_0^{+\infty}  k \bar k^{-1}(\lambda) y V(y,d) \exp(-(k +\bar k(\lambda)) y)dy=2\bar D- W_k(\lambda, \nu, d),
\label{kwc3a}
\end{equation}
where 
$W_k=I_k + J_k$
and
\begin{equation}
I_k=k^2  \int_h^{+\infty}  \bar \psi_k(y,\lambda) \bar \rho_{\bar k} V_y(y,d) dy,
\label{kwI}
\end{equation}
\begin{equation}
J_k=-k^2  \int_0^{h}  ( \psi_k(y,\lambda) \rho_{\bar k} - \bar \psi_k(y) \bar \rho_{\bar k}) V_y(y, d) dy.
\label{kwJ}
\end{equation}
Under assumption (\ref{hviscos}), $\lambda << \nu^s$ for $s \in (0,1)$ and for sufficiently large $\nu$ the term $I_k$  satisfies the estimate
\begin{equation}
|I_k| < c_1 k^2 h^{M+3} \exp(-kh) < c_2 \nu^{-4}, 
\label{kwest}
\end{equation}
which is uniform in $k$.
To estimate $J_k$ we use  the inequality 
\begin{equation}
|J_k| \le k^2  \int_0^{h}  (| (\psi_k(y,\lambda) -\bar \psi_k(y)) \rho_{\bar k}| + |\bar \psi_k(y) (\bar \rho_{\bar k}(y) -\rho_k(y))|) |V_y(y,d)| dy.
\label{kwJ2}
\end{equation}
Now we apply
relations  (\ref{psiFa1}), (\ref{rhoest1}), estimate   (\ref{psiFa}) and
$$
\sup_{y \in [0, h]}  |V_y(y, d)| <  c_3 (\kappa^{-2}  + h^{M+3}).  
$$
Then we find that  
$$
|J_k|  <  c_4 k^2 (\kappa^{-2}  + h^{M+3}) (k^2 \bar k^{-1} \exp(-\bar kh) + k \bar k^{-1} \lambda \nu^{-1})  <
$$
$$
 < c_5 (1 + \kappa^{-2}) \nu^{s-1}  
$$
for some $s \in (0,1)$ and $c_5 >0$.
 Note that in this estimate the constant $c_5$ is uniform in $k$.  
As a  result, we obtain  the  estimate
\begin{equation}
|W_k(\lambda, \nu, d)|   <  c_6 \kappa^{-2} \nu^{s-1}, 
\label{RKK}
\end{equation}
which is uniform in $k$.
Therefore, all analysis of (\ref{kwc3a}) can be made the same arguments as above
that allows us to prove the next lemma.

\begin{lemma} \label{LL}
{ Let  assumptions (\ref{hviscos}) hold,
 $N$ be a positive integer  and $K_N=\{k_1, ..., k_N \}$ be a subset of ${\bf Z}_+$ and $V(y, d)$ is defined by  
(\ref{Uy}). Moreover, let $|\lambda| < \nu^{3/4}$.  Then  in  relation (\ref{Uy}) we can choose  parameters $\kappa$ and $d$  
  such that for sufficiently large $\nu$ the roots  
 $\lambda(k, b, \nu)$ of equation (\ref{kwc3a}), which lie   in the domain $ |\lambda| < \nu^{3/4}$,  satisfy

({\bf i} )
\begin{equation}
\lambda(k, b_c, \nu)=0     \quad  k \in K_N,
\label{Spec0A}
\end{equation}
and  
\begin{equation}
  Re \ \lambda(k, b_c, \nu) < - \delta_N  \quad  k \notin K_N,
  \label{SpecA}
\end{equation}
where positive $\delta_N$ is uniform in $\nu$ as  $\nu \to \infty$;

({\bf ii}) for $k \in K_N$ the eigenvalues $ \lambda(k, b, \nu)$  change their signs at $b=b_c$ when $b$ runs the  interval $(b_1, b_2)$.   
}
\end{lemma}

{\bf Proof}.  
We repeat arguments from the proof of Lemma \ref{choiced}.  Since  $W_k(\lambda, \nu, d)$ satisfies  estimate (\ref{RKK}), we obtain new $d_j=\bar d_j(\nu)$, which are small  
perturbations of $d_j$ obtained  in Lemma \ref{choiced}. We have the estimates  
$\bar d_j(\nu) -d_j=\tilde d_j(\nu)$, where   $\tilde d_j(\nu)\to 0$ as $\nu \to +\infty$. Therefore, the $\sup |U_y(y, \nu)-  V_y(y,d)| \to 0$ as $\nu \to +\infty$. 
  Inequalities  (\ref{Spec0A}) are fulfilled that follows from equation (\ref{kwc3a})
since for sufficiently large $\nu$  estimate (\ref{RKK}) is uniform in $k$.   
$\square$.

For the case {\bf I} the assertion of the Theorem follows from this lemma.
\vspace{0.2cm}

{\em  Case {\bf II}}.

Let us consider now the second case {\bf II}.
Relations 
(\ref{psiF2}) and (\ref{rhoest1}) imply that 
$$
|\psi_k(y, \lambda)| <  c_1 \nu/|\lambda|,   \quad |\rho_{\bar k}(y) |  < c_2 \exp(-\bar k  y) |\bar k|^{-1}, 
$$
where $|\bar k| >  |\lambda|^{1/2}$.  Moreover, $|y U_y| < C h^{M+2}$. 
Thus the left hand side of eq. (\ref{kwc2}) is not more than  $R=C\nu  \log(\nu)^{M+2} k \bar |k|^{-1} |\lambda|^{-3/2}$. 
For $\nu \to +\infty$ and $|\lambda| > \nu^{3/4}$ one has $R < c\nu^{-1/8}$.  Therefore,  Eq. (\ref{kwc2}) has no solutions in the case {\bf II}, and the Theorem is proved
$\square$.

\subsection{The method of realization of vector fields}
\label{RVF}

Our next step is to show that systems (\ref{maineq1}) can exhibit a complicated large time behaviour.  For this end
we use
 the method of realization of vector fields (RVF)
proposed by
 P. Pol\'a\v cik \cite{Pol2,  Pol3},which can be described as follows.

Let us consider a family of local semiflows $S^t({ P})$ in a  Banach space $\bf H$. Suppose
 these semiflows  depend on a parameter ${ P} \in {\bf E}$, where
${\bf E}$  is another Banach space. 
Consider system (\ref{ordeq}) satisfying condition (\ref{cond1}) and (\ref{inward}).

\begin{definition}   \label{def1}
{ Let $\epsilon$ be a positive number. We say that the family of the local semiflows $S^t({
P})$ in $\bf H$  realizes the field $Q$  (dynamical system (\ref{ordeq}) 
with accuracy $\epsilon$ (briefly, $\epsilon$  - realizes) on the ball ${ B}^n(R)$, if there
exists a   parameter ${ P}={ P}(Q, \epsilon, n)$ such that

({\bf i})  the semiflow $S^t({ P})$ has a positively  invariant
 manifold ${ \mathcal M}_n(P)$ diffeomorphic to $B^n(R)$. This manifold  is defined by a map $Z: {B}^n(R) \to {\bf H}$
\begin{equation}
  z = Z(q), \quad q \in { B}^n(R), \quad z \in {\bf H}, \quad Z \in C^{1+r}
({B}^n), \label{manifRVF}
\end{equation}
where $r \ge 0$;

({\bf ii})   the restriction of the semiflow $S^t({ P})\vert_{{\mathcal M}_n(P)}$ on ${\mathcal M}_n(P)$
is defined by the system of differential equations
\begin{equation}
  \frac{dq}{dt}=Q(q) + \tilde Q(q, { P}), \quad \tilde Q \in C^{1}({ B}^n(R)),
\label{reddynam}
\end{equation}
  where
\begin{equation}
	    |\tilde Q(\cdot, { P})|_{C^1({ B}^n(R))} < \epsilon. \label{estRVF}
\end{equation}}
\end{definition}
\vspace{0.2cm}
\underline{\bf{~~~~~~~~~~~~~~~~~~~~~~~~~~~~~~~~~~~~~~~~~~~~~~~~~~~~~~~~~~~~~~~~~~~~~~~~~~~~~~~~~~~~~~~~~~~~~~~~~~}}


\end{document}